\documentclass[longauth]{aa}  

\usepackage{graphicx}
\usepackage{txfonts}
\usepackage{xcolor}
\usepackage{natbib}

\usepackage{booktabs}
\usepackage{amsmath}
\usepackage{tabularx}

\usepackage[colorlinks=true,linkcolor=blue,citecolor=magenta,pdfborder={0 0 0}]{hyperref}
\newcommand{\lya}{\relax \ifmmode {\mbox Ly}\alpha\else Ly$\alpha$\fi}
\newcommand{\ha}{\relax \ifmmode {\mbox H}\alpha\else H$\alpha$\fi}
\newcommand{\hg}{\relax \ifmmode {\mbox H}\gamma\else H$\gamma$\fi}
\newcommand{\hd}{\relax \ifmmode {\mbox H}\delta\else H$\delta$\fi}
\newcommand{\hb}{\relax \ifmmode {\mbox H}\beta\else H$\beta$\fi}
\newcommand{\sii}{\relax \ifmmode {\mbox S\,{\scshape ii}}\else S\,{\scshape ii}\fi}
\newcommand{\siii}{\relax \ifmmode {\mbox S\,{\scshape iii}}\else S\,{\scshape iii}\fi}
\newcommand{\nii}{\relax \ifmmode {\mbox N\,{\scshape ii}}\else N\,{\scshape ii}\fi}
\newcommand{\neiii}{\relax \ifmmode {\mbox Ne\,{\scshape iii}}\else Ne\,{\scshape iii}\fi}
\newcommand{\oii}{\relax \ifmmode {\mbox O\,{\scshape ii}}\else O\,{\scshape ii}\fi}
\newcommand{\oi}{\relax \ifmmode {\mbox O\,{\scshape i}}\else O\,{\scshape i}\fi}
\newcommand{\oiii}{\relax \ifmmode {\mbox O\,{\scshape iii}}\else O\,{\scshape iii}\fi}
\newcommand{\hii}{\relax \ifmmode {\mbox H\,{\scshape ii}}\else H\,{\scshape ii}\fi}
\newcommand{\heii}{\relax \ifmmode {\mbox He\,{\scshape ii}}\else He\,{\scshape ii}\fi}
\newcommand{\hei}{\relax \ifmmode {\mbox He\,{\scshape i}}\else He\,{\scshape i}\fi}
\newcommand{\hi}{\relax \ifmmode {\mbox H\,{\scshape ii}}\else H\,{\scshape i}\fi}
\newcommand{\ciii}{\relax \ifmmode {\mbox C\,{\scshape iii}}\else C\,{\scshape iii}\fi}
\newcommand{\cii}{\relax \ifmmode {\mbox C\,{\scshape ii}}\else C\,{\scshape ii}\fi}
\newcommand{\ariii}{\relax \ifmmode {\mbox Ar\,{\scshape iii}}\else Ar\,{\scshape iii}\fi}
\newcommand{\ariv}{\relax \ifmmode {\mbox Ar\,{\scshape iv}}\else Ar\,{\scshape iv}\fi}
\newcommand{\feii}{\relax \ifmmode {\mbox Fe\,{\scshape ii}}\else Fe\,{\scshape ii}\fi}
\newcommand{\Mgii}{\relax \ifmmode {\mbox Mg\,{\scshape ii}}\else Mg\,{\scshape ii}\fi}
\newcommand{\nev}{\relax \ifmmode {\mbox Ne\,{\scshape v}}\else Ne\,{\scshape v}\fi}
\graphicspath{{./}{figures/}}

\begin{document} 
    \title{The k-MENDEL sample of local analogs to reionization galaxies}
   
   \titlerunning{EELG characterization in DESI}
\authorrunning{Bonatto \& Amorín et al.}
   \subtitle{Spectral identification of EELGs and properties of green peas in DESI}   
\author{L.~Bonatto\inst{\ref{inst1},\ref{inst2}}\thanks{Corresponding authors: L. Bonatto, \email{albonatto@gmail.com}; R. Amorín, \email{amorin@iaa.es}} 
\and
R.~Amor\'in\inst{\ref{inst3}}
\and
A. Giménez-Alcázar\inst{\ref{inst3}} 
\and
J.A. Fernández-Ontiveros\inst{\ref{inst4},\ref{inst5}} 
\and
A. Hernán-Caballero\inst{\ref{inst4},\ref{inst5}}
\and
S. Suárez\inst{\ref{inst6}}
\and
J.M. Vílchez\inst{\ref{inst3}} 
\and
E. Pérez-Montero\inst{\ref{inst3}} 
\and
M. Llerena\inst{\ref{inst7}} 
\and
J. Sánchez Almeida\inst{\ref{inst8},\ref{inst9}}
}
    
    \institute{
    Vicerrectoría de Investigación y Postgrado, Universidad de La Serena, 1700000, Chile \label{inst1}
    \and
    Istituto Nazionale di Geofisica e Vulcanologia, Via di Vigna Murata 605, 00143, Rome, Italy \label{inst2}   
    \and
    Instituto de Astrof\'{i}sica de Andaluc\'{i}a (CSIC), Apartado 3004, 18080 Granada, Spain \label{inst3}
    \and
    Centro de Estudios de F\'{\i}sica del Cosmos de Aragón (CEFCA), Unidad Asociada al CSIC, Plaza San Juan 1, E--44001 Teruel, Spain \label{inst4}
    \and
    Unidad Asociada CEFCA--IAA, CEFCA, Unidad Asociada al CSIC por el IAA y el IFCA, Plaza San Juan 1, 44001 Teruel, Spain \label{inst5}
    \and
    Departamento de Astronomía, Universidad de La Serena, Avenida Raúl Bitrán 1305, La Serena, Chile \label{inst6}
    \and
    INAF - Osservatorio Astronomico di Roma, Via di Frascati 33, 00078, Monte Porzio Catone, Italy \label{inst7}
    \and
    Instituto de Astrof\'{i}sica de Canarias, La Laguna, Tenerife, E-38200, Spain  \label{inst8}
    \and
    Departamento de Astrof\'\i sica, Universidad de La Laguna, Tenerife, Spain
    \label{inst9}
    }
    
   \date{Received ----; accepted ----}

 
  \abstract
{Low-mass galaxies undergoing intense starburst episodes exhibit spectra dominated by extreme nebular emission and faint stellar continua. These extreme emission-line galaxies (EELGs) provide key laboratories to study massive star formation, chemical enrichment, feedback, and the escape of ionizing photons in low-metallicity environments.}
{We exploit the large spectroscopic volume of the DESI survey to assemble the k-Means of Extreme Nebulae from DEsi outLiers (k-MENDEL), a homogeneous and statistically robust sample of EELGs at $z<1$ selected through their unusual spectral properties. Our goal is to characterize their physical conditions and scaling relations, and to assess their role as local analogs of galaxies in the early Universe.}
{Using an automatic spectroscopic k-means (ASK) classification, we identify $\sim$16\,000 EELGs at $0.01<z<0.96$. Stellar masses, star-formation rates, and dust attenuation are derived from spectrophotometric SED fitting, while emission-line measurements provide nebular extinction, ionization diagnostics, and $T_e$-based gas-phase metallicities.}
{k-MENDEL charcaterizes a variety of EELGs, including blueberry and green pea galaxies. It spans stellar masses from $10^{6}$ to $10^{10}$\,M$_{\odot}$, star-formation rates of $\sim$0.1-100\,M$_{\odot}$\,yr$^{-1}$, and gas-phase metallicities of $12+\log(\mathrm{O/H})\sim7.0$-8.5, extending previous SDSS samples toward higher redshift and lower masses and metallicities. 
EELGs lie systematically above the star-forming main sequence, with the highest EW systems reaching specific SFR (sSFR) up to  $\sim$100 Gyr$^{-1}$. 
EELGs follow a shallower mass-metallicity relation offset by $\sim$0.3-0.5 dex from local relations, closely resembling   the locus of young galaxies observed with \textit{JWST} at $z>3-10$. 
Projecting along the fundamental metallicity relation does not reduce the large intrinsic metallicity scatter, indicating strong departures from stationary or "bathub" model predictions and suggesting massive inflows of metal-poor gas sustaining star formation and eventually leading to strong feedback. 
A small fraction ($\sim$6\%) of objects show AGN-like ionization signatures, while the most extreme star-forming systems reach high ionization conditions ($O_{32}>5$-60) comparable to confirmed low-redshift Lyman-continuum emitters and reionization-era JWST galaxies. }
{These results support the interpretation of EELGs as short-lived, highly non-equilibrium phases in the evolution of low-mass galaxies and highlight their importance as nearby analogs of galaxies likely driving cosmic reionization.}

   \maketitle
%

%
\section{Introduction} \label{sec:intro}

Nebular emission lines are among the most powerful tracers of star formation and nuclear activity in galaxies. In the optical regime, the most prominent features are the [\oiii]$\lambda\lambda$4959,5007 doublet and the Balmer lines, particularly H$\alpha$. Their large equivalent widths (EWs) provide a direct and easily measurable signature of intense starbursts and active galactic nuclei (AGNs), often recurrent phases of galaxy evolution \citep{Kewley2019}.  
  
At the high-end of the EW distribution lies the population of \textit{extreme emission-line galaxies} (EELGs), generally defined by rest-frame [\oiii]$\lambda$5007\AA\ EWs ranging $\sim$100-2000\,\AA, reflecting ongoing bursts of star formation. These are typically compact, low-mass (M$_{\star}<$\,10$^{10}$\,M$_{\odot}$) systems undergoing very young and vigorous starbursts, sometimes accompanied by an AGN \citep[e.g.][]{Amorin2015,EPM2021}. They are found across cosmic time, from the nearby Universe to the epoch of reionization.

At low redshift ($z<1$), pioneering objective-prism and spectroscopic observations \citep[e.g.][]{SargentSearle1970,Terlevich1991} identified H\,{\sc ii} galaxies through their strong emission lines. Larger spectroscopic surveys, such as the Sloan Digital Sky Survey (SDSS), have since provided samples of hundreds to a few thousand EELGs that allowed detailed studies of their physical and chemical properties, with different samples being nicknamed according to their defining properties and redshift range, such as blueberries at $z<0.1$ and green peas (GP) at $0.1<z<0.4$ \citep[e.g.][]{Kniazev2004, Cardamone2009, Amorin2010, Amorin2012, Izotov2011, Yang2017_bb, EPM2021, Lumbreras-Calle2022, delmoral2024}. Deeper spectroscopic campaigns in cosmological fields later revealed similar samples of EELGs up to $z\sim1$ \citep[e.g.][]{Atek2011,Amorin2014,Amorin2015,Ly2014,Calabro2017,delmoral2024}, showing that these systems share many characteristics with higher-redshift galaxies, namely compact morphologies ($r_{50}\lesssim1$\,kpc), very high specific star formation rates (sSFR~$\gtrsim10^{-8}\,\text{yr}^{-1}$), low metallicities ($12+\log(\text{O/H})\sim7.0$–8.4), and extreme ionization conditions. 

Although rare in the local Universe, their number density increases rapidly with redshift \citep[e.g.][]{Maseda2018,Atek2024Natur,Khostovan2024}. Recent \textit{JWST} discoveries demonstrate that EELGs are key analogs of primeval galaxies, providing a direct window into the physical processes that dominated the first billion years, reionization, rapid star formation, and early chemical enrichment \citep[e.g.][]{Schaerer2022, Rhoads2023, Arellano2022,Trump2023}. 
Beyond $z>2$, observations with the \textit{Hubble Space Telescope} (HST) identified samples of compact EELGs dominated by [\oiii] emission \citep{vanderWel2011,Maseda2018,Tang2019,Du2020,Llerena2023}. The advent of infrared telescopes, first Spitzer and more recently the \textit{James Webb Space Telescope} (JWST) has greatly expanded these samples to higher redshifts by exploiting infrared color excesses from strong emission lines \citep{Smit2014,Faisst2016,Castellano2017,Hutchison2019} and detailed SED modeling  \citep{Llerena2024,Boyett2024,Davis2024,Endsley2024}. Spectroscopic observations with \textit{JWST} 
are now providing unprecedented access to the physical conditions of EELGs at the highest redshifts, well into the era of reionization \citep[EoR, e.g.][]{Arrabal2023Natur,Castellano2024,Carniani2024}.  

Simulations and observations indicate that bursty star formation is a prevalent mode of growth in low-mass galaxies at high redshifts \citep[e.g.,][]{Shen2014,Sparre2017,Atek2022,Simmonds2025}. Observationally, such galaxies exhibit high ionizing photon production efficiencies at all redshifts \citep[e.g.][]{Chevallard2018,Flury2022a,Endsley2023, Simmonds2025, Pahl2025, Llerena2025, Gimenez-Alcazar2025} and frequently are found as strong Ly$\alpha$ emitters (LAEs; \citealt{Cowie2011,Henry2015,Berg2022,Izotov2024_lya}), although not all EELGs show escaping Ly$\alpha$ or Lyman continuum (LyC) radiation. At $z\sim0.3$, deep HST spectroscopy has revealed that a subset of EELGs emit significant LyC flux \citep[e.g.][]{Izotov2016,Izotov2018a,Flury2022a,Flury2022b}, making them unique laboratories for studying the production and escape of ionizing photons, key ingredients for understanding the galaxies that reionized the Universe. 

Despite recent progress, several key questions about the nature of EELGs at different redshifts are still uncertain, such as their contribution to the cosmic SFR density \citep[e.g.][]{Khostovan2024}, their nebular excitation and star formation and chemical enrichment histories \citep[e.g.][]{Amorin2010,Calabro2017,Arellano2022,Curti2024_mzr,Sanders2024_auroral,Llerena2024,Llerena2026, Boyett2024,Katz2025}, and the physical mechanisms regulating ionizing-photon leakage \citep[e.g.][for a review]{Jaskot2025}. Current samples remain limited in size, redshift coverage, and homogeneity, hindering statistically robust tests of how bursty star formation shapes galaxy scaling relations, and ionizing output across cosmic time.  
Together, these open issues point to an incomplete understanding of how EELGs fit within the broader framework of galaxy growth and chemical evolution.

Addressing these issues requires large, deep, and homogeneous spectroscopic datasets that probe a wide dynamic range in redshift and luminosity. In particular, providing new light into the physical link between local analogs with their high-redshift counterparts, a more complete census of EELGs at $z<1$ becomes essential to address statistical studies. The largest previous spectroscopic studies on EELG to date are based on a few  thousand objects out to $z\lesssim0.5$ from SDSS \citep{EPM2021} or LAMOST surveys \citep{Liu2023}. These samples are typically limited to the brightest EELGs, especially at $z>0.1$, biasing studies against the least massive and more metal-poor EELGs (stellar masses $\log$($M_{\star}/M_{\odot})<$\,8 and $12+\log(O/H)<$7.6), which include the most extreme, possibly nearest analogs of the dominant population within the reionization era.   

The \textit{Dark Energy Spectroscopic Instrument} \citep[DESI;][]{DESI2022} provides such an opportunity, delivering spectra for millions of galaxies across a broad redshift interval. This work presents the first large, statistically significant sample of EELGs from DESI’s Early Data Release, enabling the characterization of their emission-line and physical properties up to $z\sim1$ and allowing us to place EELGs within the broader context of galaxy scaling relations. The present paper lays the groundwork for a series of studies exploring the scaling relations, ionization diagnostics, and chemical abundance calibrations of these extreme systems. As a demonstrator, this work will provide the basis for finding even larger samples with future DESI data releases and calibrate more sophisticated methods to search for EELGs in wide spectro-photometric surveys out to $z\sim2$ \citep[e.g.][]{Gimenez-Alcazar2025}.
Moreover, this study presents and discuss the basic characterization and physical properties of EELGs based on an the largest spectroscopic sample compiled to date. 

The paper is organized as follows. Section~\ref{sec:data} describes the data and sample selection. Section ~\ref{sec:ask} presents the methodology used to identify EELG candidates. Sections~\ref{sec:methods} and ~\ref{sec:results} present the emission-line analysis and derivation of physical quantities such as star formation rates, metallicities, and ionization parameters. Section~\ref{sec:discussion} discusses the EELG population within the DESI sample and their role as local analogs of high-redshift galaxies. Section~\ref{sec:summary} summarizes the main results. We adopt a flat $\Lambda$CDM cosmology with $H_0=70\,\mathrm{km\,s^{-1}\,Mpc^{-1}}$, $\Omega_m=0.3$, and $\Omega_\Lambda=0.7$.

\section{Data and parent sample} \label{sec:data}

\label{sec:ask}
\begin{figure*}[!t]
    \centering
\includegraphics[width=0.99\textwidth]{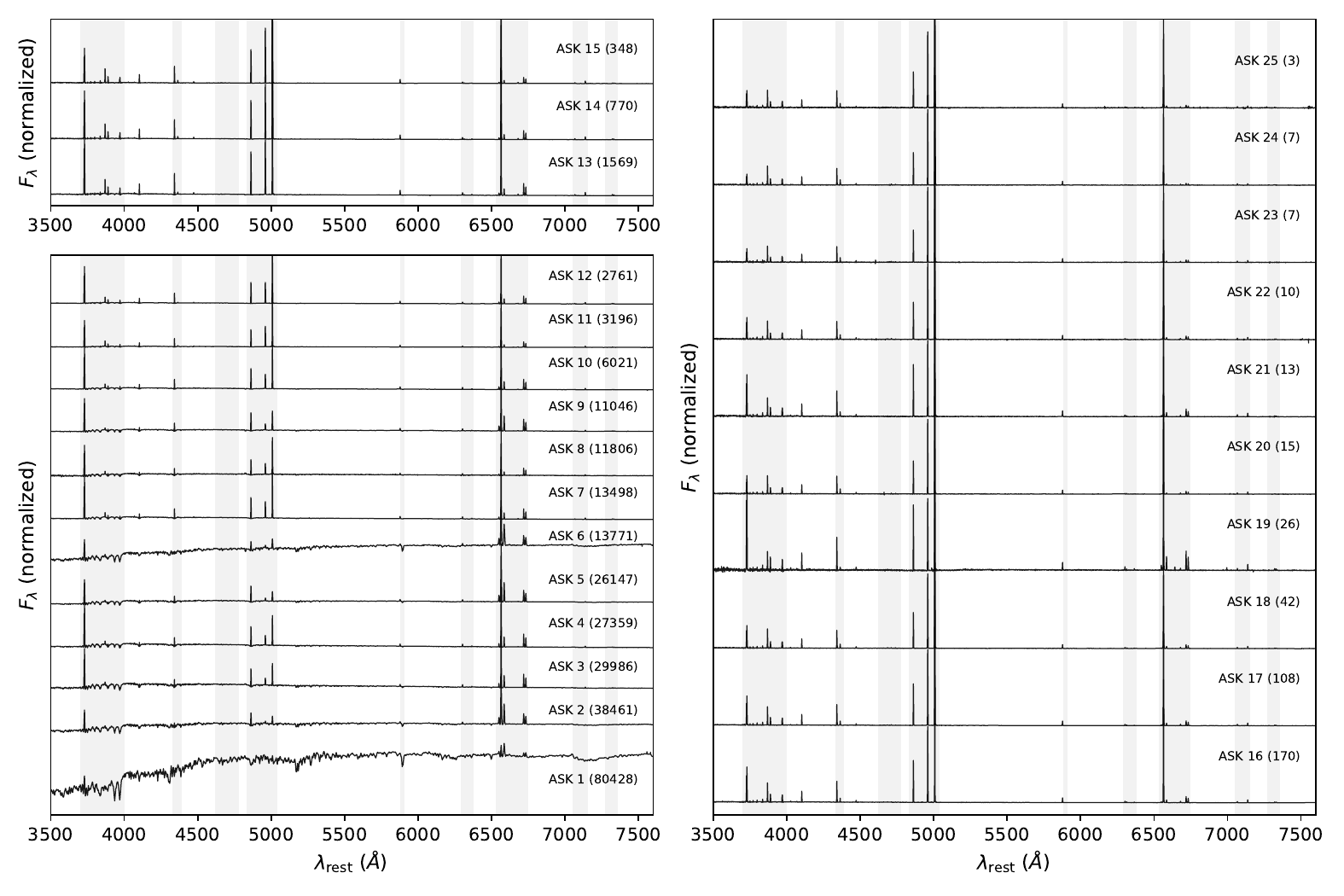}
    \caption{(\textit{Left}) Rest-frame optical median stack spectra of the major (bottom) and minor (top) ASK classes. 
    (\textit{Right}) Median stack spectra of galaxies identified as \textit{outliers} (see text for details). 
    All spectra are normalized to the continuum level at $\lambda$\,4800\,\AA\ and vertically shifted for clarity. 
    Grey shaded regions indicate the spectral windows used for the ASK classification. 
    Labels denote the number of galaxies contributing to each composite spectrum. 
    A zoom-in of the blue wavelength range ($3800$--$4800$\,\AA) is shown in Fig.~\ref{fig:outliers}.
    }
    \label{fig:clusters}
\end{figure*}

\subsection{DESI EDR spectroscopy}
We use optical spectroscopy from the Dark Energy Spectroscopic Instrument (DESI) cosmological survey \citep{DESI2016a,DESI2016b,DESI2024a,DESI2024b}, which provides homogeneous medium-resolution spectra ($R\sim2000$--5000) over the wavelength range 3600--9800\,\AA\ for millions of extragalactic targets selected from the DESI Legacy Imaging Surveys \citep{Zou2017,Dey2019}. Observations are carried out with a 5000-fiber multi-object spectrograph mounted at the prime focus of the Mayall 4\,m telescope at Kitt Peak \citep{DESI2022,Miller2024,Silber2023}. The wide spectral coverage, large multiplexing capability, and survey depth makes DESI particularly well suited to identify rare populations of extreme emission-line galaxies over large cosmic volumes.

In this work, we use public spectra from the Early Data Release (EDR), obtained during Science Verification observations and covering 732 tiles (the Fuji data; \citealt{DESI2024b}). The data reduction and spectrophotometric calibration are described by \citet{Guy2023}, while value-added catalogs provide spectral classification and precise redshift measurements from the \textsc{Redrock} pipeline \citep{Brodzeller2023}. 

From the $\sim$2 million spectra in the EDR redshift catalog, we selected non-duplicated objects (\texttt{MAIN-SPECTRA}) classified as galaxies (\texttt{SPECTYPE = GALAXY}) with reliable redshift solutions (\texttt{ZWARN = 0} and \texttt{DELTACHI2 > 40}). We further restricted the sample to $0.01 \leq z \leq 0.96$, ensuring that the \hb\ line and the [\oiii]$\lambda\lambda$4959,5007 doublet, key tracers of ionization and gas-phase metallicity, are always covered by the DESI spectral range. These criteria yield a parent sample of 820\,240 unique galaxy spectra that forms the basis for the automated identification of EELGs described below.

\section{Methodology}

We performed an automatic unsupervised classification of DESI EDR galaxy spectra using the ASK method of \citet{SA2010}, which is based on a small set of physically motivated spectral features. This approach has been successfully employed by \citet{Morales-Luis2011} to select extremely metal-poor galaxies and by \citet{EPM2021} to identify EELGs in SDSS-DR7 spectra. 
The primary goal is not to provide a detailed spectral taxonomy of galaxies, but to develop an efficient tool to identify EELG candidates. This method assumes that EELG spectra are intrinsically rare \citep[e.g.,][]{SA2010,SanchezAlmeida2012} and avoids relying on pre-computed emission-line measurements for the entire survey. As we demonstrate below, it provides a powerful and scalable way to isolate specific spectral classes such as EELGs, particularly in view of the increasingly large future DESI data releases.

In the following, we briefly summarize the ASK method, describe its implementation on DESI spectral data, and explain the procedure adopted to identify EELG candidates.

\begin{figure*}[!t]
    \centering
\includegraphics[width=0.33\textwidth]{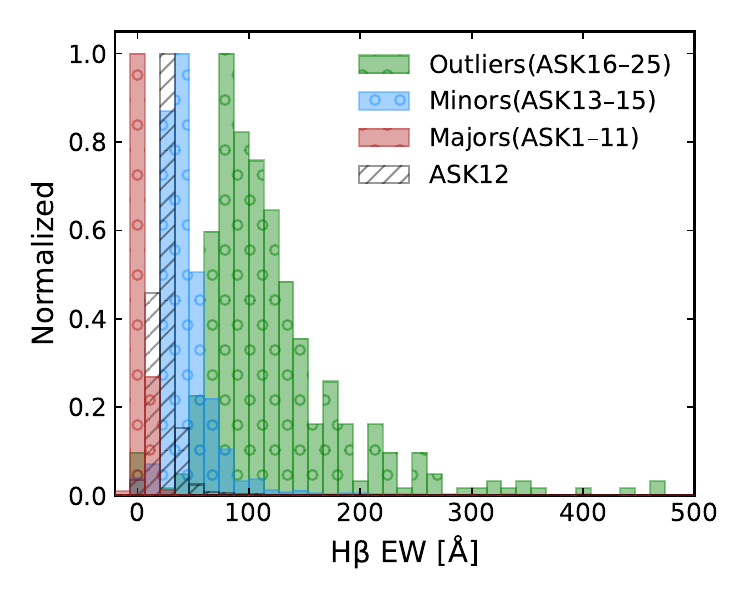}
\includegraphics[width=0.33\textwidth]{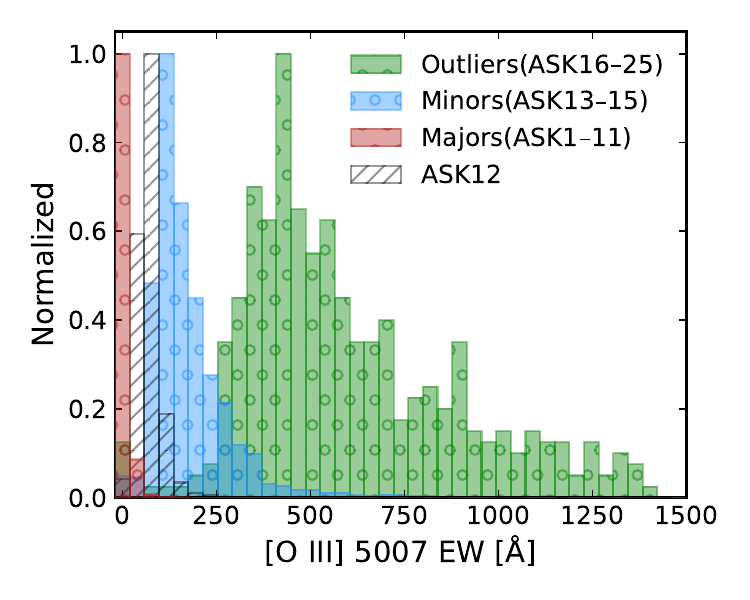}
\includegraphics[width=0.33\textwidth]{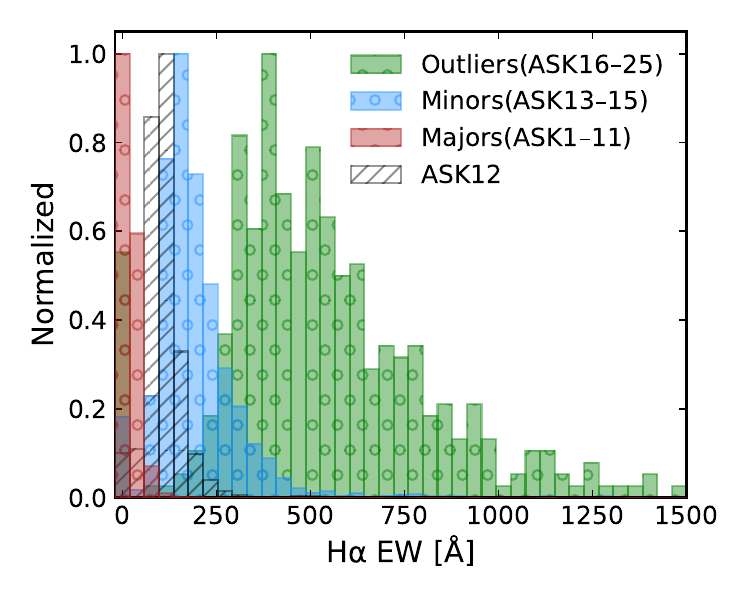}
    \caption{Distributions of emission-line equivalent widths for galaxies classified in major (ASK1-12) and minor (ASK13-15) ASK classes. 
    For comparison, we also show galaxies belonging to ASK12, which represents the most extreme major class, and spectral outliers (ASK16-25). 
    Equivalent widths are taken from the DESI-EDR value-added catalog of \citet{Zou2024}. 
    Minor classes and outliers are clearly associated with progressively larger equivalent widths, illustrating that the ASK classification efficiently isolates galaxies with extreme nebular emission.
    }
    \label{fig:hist_ask}
\end{figure*}

\subsection{ASK classification method}
ASK (Automatic Spectroscopic K-means) \citep{SA2010} is an unsupervised k-means-based classification framework that groups galaxies according to similarities in their spectra.

The k-means clustering algorithm iteratively assigns data points (here, galaxy spectra) to clusters according to their similarity to the cluster centers. 
Each spectrum is represented as a vector in a high-dimensional space, where each dimension corresponds to a rest-frame wavelength. 
The algorithm starts by selecting $k$ initial cluster centers (typically at random), assigns each spectrum to the nearest center using a distance metric (usually Euclidean), and then updates the centers by computing the mean spectrum of the assigned members. 
These assignment and update steps are repeated until convergence. 
Convergence occurs when no spectrum is re-assigned to a different cluster in consecutive iterations. 
The number of clusters $k$ is chosen a priori. 
The final output consists of the cluster centers and the classification of all spectra.

The ASK method \citep{SA2010} replaces the random initialization of cluster centers with predefined initial clusters that naturally emerge from the dataset. It begins by randomly selecting a small set of candidate centers (typically around 10) and performing a single k-means iteration. 
The cluster containing the largest number of spectra is retained as an initial center, its members are removed from the sample, and the procedure is repeated until all galaxies are assigned to predefined initial clusters.
Using these centers, the ASK method performs a standard iterative k-means classification with an additional convergence criterion: the loop stops when the classifications between successive iterations differ by less than 0.01\%; i.e., 99.99\% of the assignations do not vary between two iterations.

To reduce computational cost in the high-dimensional spectral space, ASK relies on selected bandpasses centered on key spectral features, thereby limiting the number of wavelengths used in the classification.

\subsection{Implementation of ASK on DESI galaxy spectra }

In practice, we implemented the ASK procedure using the \textsl{k}-means algorithm as provided by the \textsl{scikit-learn} library \citep{scikit-learn}. 
We introduced a modification that intuitively provides better initial clusters: we conduct the first iteration ten times and retain the best classification, characterized by the lowest inertia (i.e., the sum of variances within each cluster). This adjustment is accomplished by setting n$\_$init $=10$ in the k-means method.
The pre-processing of all galaxy spectra and the main steps of the classification are outlined in Fig.~\ref{fig:proc-steps} and described in larger detail in Section~\ref{app:ask}.

The spectral data for the k-means classification must share the same rest-frame wavelength range. 
However, the DESI dataset samples a wide range of redshifts, limiting the wavelength range to perform the classification. 
Thus, we divided our dataset into two subsets: a low redshift set (267,568 spectra with $0.01\leq z \leq 0.25$) and a high redshift set (552,676 spectra with $0.25 < z < 0.96$). 
We performed the classification over the low-redshift set. 
This choice dictates the wavelength range to perform the classification: at z $<$ 0.25, the spectra span a wavelength range between 3500 and 7500\AA, thus including the Balmer break and the main optical nebular emission lines indicative of star-forming galaxies and AGNs. 
To reduce the number of wavelengths included in the classification, we defined a set of spectral windows for the ASK analysis. 
These windows are centered on key nebular emission features, as well as on the Balmer and D4000 breaks (see Table~\ref{tab:windows}). 

Finally, the classification of the complete parent sample (low-redshift and high-redshift subsets) was performed through quality assignment following \citet{SA2010}. 
A quality value was defined to assess the membership of each galaxy to the classes identified by the k-means algorithm.
For each cluster, we computed the weighted Euclidean distances between the spectra of the low-redshift subset used to derive the k-means classification and their corresponding cluster center, considering only the wavelength windows adopted in the analysis. 
These distances define an empirical distribution for each cluster, from which we derived a cumulative probability function (CDF) giving the probability of finding a spectrum in that cluster with a distance equal to or larger than a given value.
For each cluster, the observed CDF was fit with an appropriate quality function (QF) using a non-linear least-squares method. This QF, calibrated exclusively on the low-redshift classification sample, was then evaluated for every spectrum in the complete parent sample.
The quality of a spectrum with respect to a class is defined as the value of the corresponding cluster QF evaluated at its distance to that cluster center. 
High-quality values indicate a strong association with a class, whereas very low best-quality values identify spectra poorly represented by any cluster (i.e., potential outliers). 
Quality values also provide a measure of how well-defined each class is in the classification space (see Section~\ref{app:ask}).

\subsection{Selection criteria for EELG candidates at $z\le0.25$}
Using the ASK classification and quality framework described above, we now identify the spectral classes hosting EELG candidates. 
As mentioned in Section~3.2, the full set of galaxies was split into two subsets. 
The ASK classification was applied to the low redshift subset consisting of 267,568 galaxy spectra. 
This analysis yielded 25 distinct spectral classes (named as ASK1 to ASK25), whose centroids are shown in Fig. \ref{fig:clusters}.  

Based on the population size of each ASK class and the ability to obtain reliable QF (i.e., stable fits of the CDF using sufficient galaxies and centroid distances), we defined two groups, Major and Minor classes. 
Major classes (ASK1–ASK12) contain 264,480 galaxies (98.85\% of the total), and Minor classes (ASK13–ASK15) comprise 2,687 galaxies (1.0\%). 
Both groups have classes with well-defined quality functions. In contrast, ASK16–ASK25 (401 galaxies, 0.15\%), which we defined as \textit{outliers}, lack QF because their small populations and limited distance distributions do not allow for a robust fit.

This ASK classification provides a practical way to identify EELG spectra.
Instead of measuring the entire EDR and selecting EELGs by their EWs, as is typically done in smaller surveys, here we measured the 25 ASK centroids and assigned our preferred EELG definition to these ASK classes. 
In Table~\ref{tab:ASK-EW}, we show the EW measurements for each ASK class from ASK10 to ASK25. Classes from ASK1 to ASK9 are not reported as they all have smaller EWs. 
From Table~\ref{tab:ASK-EW} we see that all ASK classes from ASK13 to ASK25 (i.e. minor classes and outliers) have average EWs within the typical cuts used for EELG  definition in spectroscopic surveys EW([\oiii]$_{5007}$)$>$\,100-200\AA\ \citep[e.g.][]{Atek2011, vanderWel2011, Amorin2015, Llerena2024, Boyett2024}. 
The outliers are ASK classes with the largest EW in the parent sample, including the few SDSS EELGs (i.e. green peas and blueberry galaxies) studied in previous works that have DESI spectra. 
Therefore, within our ASK classification method, one operative way to define EELGs is based on those galaxies with spectra classified in these minor and outlier ASK classes. 
Importantly, the ASK classes hosting the EELG candidates are also those that are best defined in the classification space (Section ~\ref{app:ask}). 
This emphasizes the strength of the algorithm in separating galaxies with spectra that are clearly distinct from the bulk of the population.

\begin{figure}[!t]
    \centering
    \includegraphics[width=0.5\textwidth]{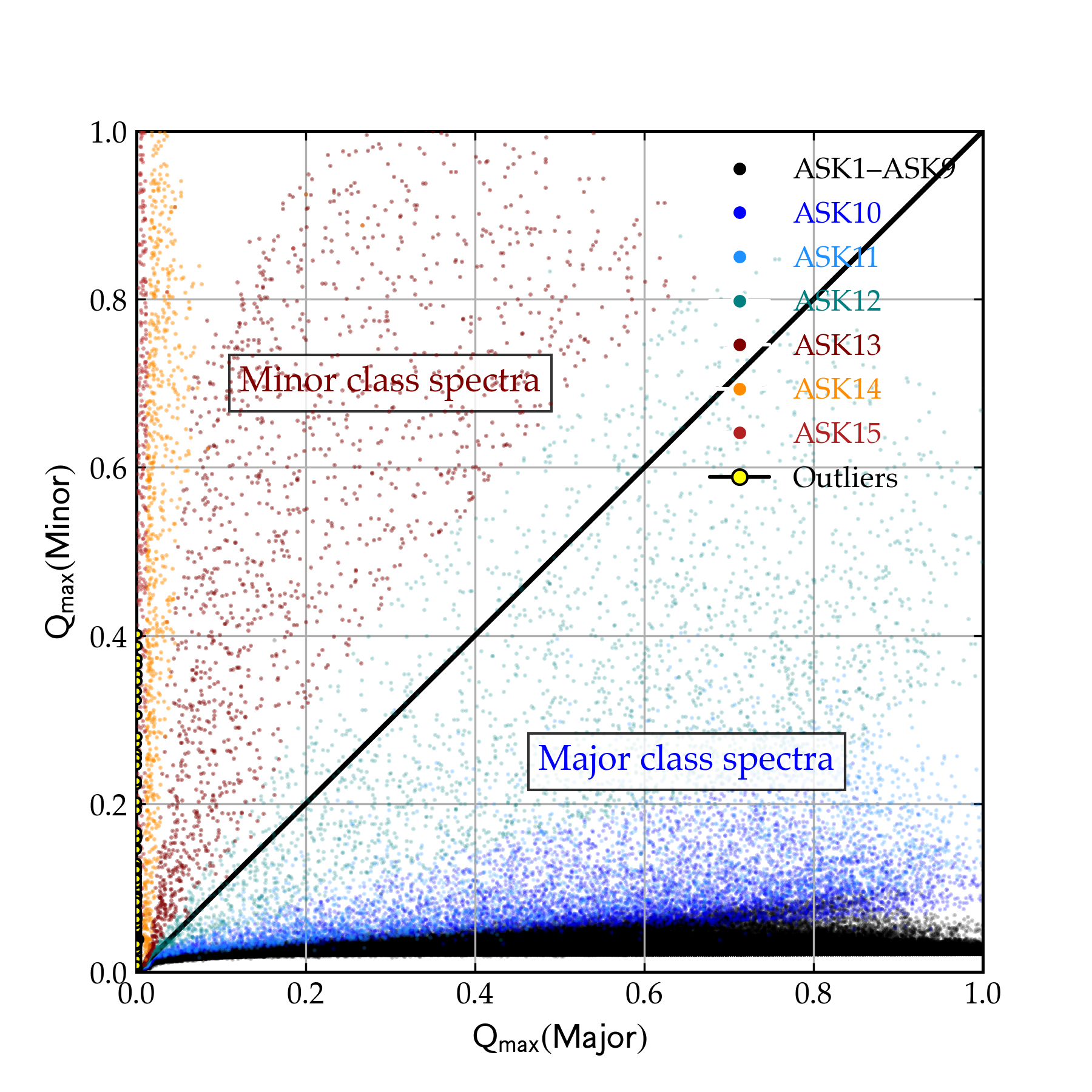}
    \caption{Maximum quality associated with major and minor ASK classes for galaxies in the low-redshift subsample used to define the ASK classification ($0.01<z\leq0.25$). 
    Each point represents one galaxy, showing the highest quality value among major classes (ASK1-12), $Q_{\rm max}$(Major), versus that among minor classes (ASK13-15), $Q_{\rm max}$(Minor). 
    Colours indicate the ASK class providing the best spectral match, including outliers (ASK$\geq$16). 
    The solid black line marks the one-to-one relation. 
    Galaxies located above this demarcation are preferentially associated with rare or extreme spectral classes and are therefore selected as EELG candidates.
    }
    \label{fig:Qmax}
\end{figure}

\subsection{Validation of EELG selection criteria}

We tested the above EELG candidate selection on the $z\leq$\,0.25 subsample, against individual measurements of bright emission line EWs of spectra that belong to ASK classes using the DESI VAC of \citet{Zou2024}. In Fig.~\ref{fig:hist_ask}, we show the EW distribution for \hb, [\oiii]$_{5007}$, and \ha, of galaxies in major, minor, and outlier classes. We only used the subset of galaxies classified by k-means with good EW measurements (errors below 50\%). Among the major classes, we identify galaxies in ASK12 to illustrate how these classes form a continuous distribution in EW. From Fig.~\ref{fig:hist_ask}, we demonstrate that a very small fraction of selected outliers ($\sim$1.3\% ) and minor (ASK13-ASK15) classes ($\sim$13\%) have EW$_{5007}<$\,100\AA, and that only $\sim$10\% (0.8\%) of galaxies within major and minor classes not included in our selection have EW$_{5007}\sim$\,100-200\AA ($>$200\AA). It is interesting to notice that galaxies with the largest EW$_{5007}>$\,500\AA\ are only identified as outliers in our ASK classification. We therefore conclude that our method can naturally identify EELGs with cuts in  EW$_{5007}$ from $\gtrsim$\,100\AA\ by selecting candidates with maximum quality assignation to minor (ASK13-ASK15) and outlier (ASK16-ASK25) classes. 

Using the major and minor classes with defined QF (i.e. ASK1–ASK15), we calculated quality values for the galaxies in the low-redshift subset already classified by k-means. This provides each galaxy with a full set of quality values associated with the ASK1–ASK15 classes. 
From these values, we computed for each galaxy the maximum quality associated with a major (ASK1 to ASK12) and minor classes (ASK13 to ASK15), $Q_{max}$(Major) and $Q_{max}$(Minor), respectively.  In Fig.~\ref{fig:Qmax}, we show the distribution of spectra belonging to the ASK classes. We note that spectra within major classes (ASK1 to ASK9, black points) show very low $Q_{max}$(Minor), while spectra with large maximum quality associated with a minor class have very small $Q_{max}$(Major). As shown in Fig.~\ref{fig:hist_ask}, we identified some ASK classes, particularly ASK10, ASK11, and ASK12, that may include galaxies with similar spectral features, thus overlapping in the EW distribution. This is also evidenced in Fig.~\ref{fig:Qmax}, where spectra increase their $Q_{max}$(Minor) values from ASK10 to ASK12, with values above the identity and reaching the lower boundary of ASK13, the first minor ASK class. This means that for galaxies in these ASK classes, a selection criterion based on $Q_{max}$(Major) $< Q_{max}$(Minor) (i.e. spectra above the identity in Fig.~\ref{fig:Qmax}) may result in the inclusion of some galaxies with spectral features that are consistent with these major classes. 
Outliers, in turn, are  characterized by $Q_{max}$(Major) $\sim$ 0 and $Q_{max}$(Minor)$< 0.4$.

\subsection{The k-MENDEL EELG sample at 0.01\,$<z<$\,0.96}

Once our selection criteria are tested and validated in the low-redshift ASK subset, we performed the class designation for the remaining sample of 556819 spectra in the DESI EDR parent sample. 
We derived a quality value with respect to each ASK centroid, representing the likelihood that the galaxy belongs to a given class, and computed $Q_{max}$(Major) and $Q_{max}$(Minor). We find that 544043 or 97.7\% of them have the largest quality consistent with major classes (i.e. ASK1 to ASK12); 10538 or 1.9\% have qualities consistent with minor classes (i.e. ASK13 to ASK15), and 2238 or 0.4\% have maximum quality values equal to zero and are thus considered as outliers. From the latter, we find that 348 spectra have poor quality according to the criteria in \cite{Lan2023} (e.g. VI 0, 1, and 2) or correspond to objects with all three key emission lines (\hb\ and \oiii) affected by strong telluric contamination, so we removed these objects for the subsequent analysis. 
Based on the $Q_{max}$(Major)-$Q_{max}$(Minor) diagram (Fig \ref{fig:Qmax}), we then select only galaxies that have spectra with $Q_{max}$(Major) $\le Q_{max}$(Minor). 

\begin{figure*}[!t]
    \centering
\includegraphics[width=0.95\textwidth]{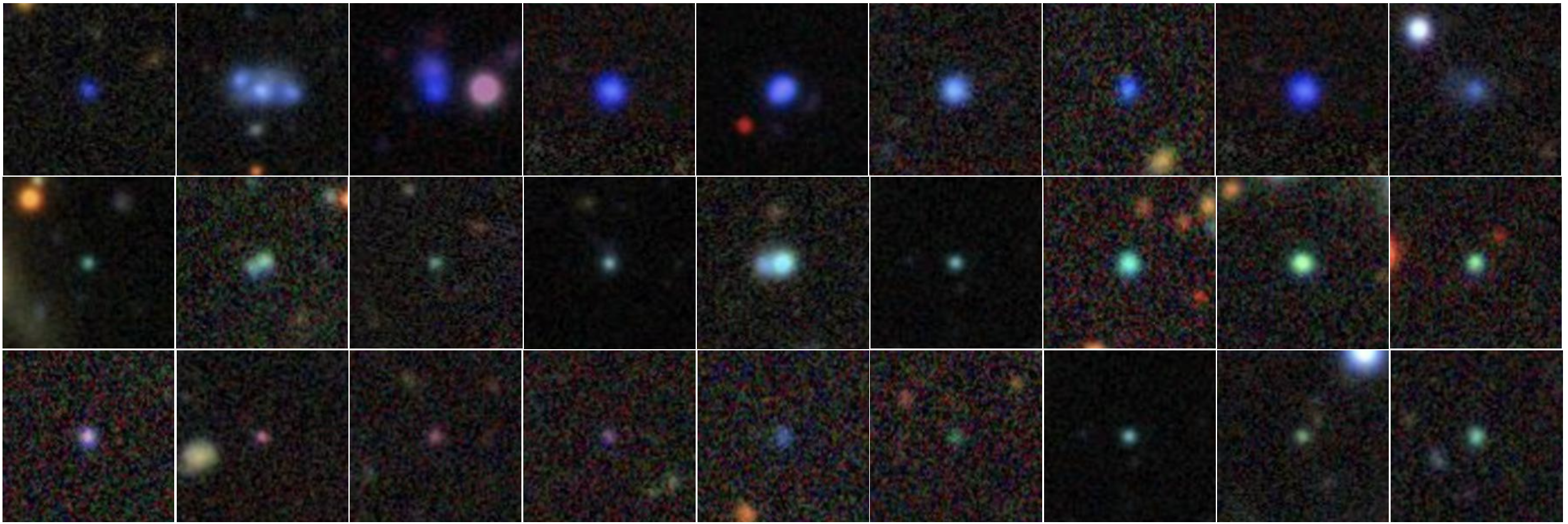}
    \caption{Examples of DESI EELGs from the k-MENDEL sample. Legacy Survey DR10 color-composite images are shown for a representative subsample spanning a range of stellar masses, redshifts, and emission-line equivalent widths. Images are displayed with North up and East to the left, and each panel corresponds to a field of $20\arcsec \times 20\arcsec$. Most systems appear compact or nearly unresolved at DESI imaging resolution, consistent with intense star formation occurring in low-mass galaxies. In some cases, multiple bright clumps or disturbed morphologies are visible, suggestive of clumpy star formation or possible interactions.}
    \label{fig:cutouts}
\end{figure*}

Using the above criteria, the final EELG sample comprises 15,014 EELG candidates at $0.01 < z < 0.96$, which we refer to as the \textit{k-MENDEL EELG sample}.
In Fig.~\ref{fig:cutouts}, we illustrate the morphology of a representative subset of selected galaxies at different redshifts, using RGB color images. Overall, our EELGs are compact galaxies with faint continuum and bright nebular emission, which, according to redshift, gives the colored appearance in the RGB images. Thus, our sample includes the so-called blueberry and green pea galaxies at $z\lesssim$\,0.1 and 0.1\,$\lesssim z \lesssim$\,0.35, respectively, along with reddish EELGs at larger redshifts. According to the DESI-EDR VAC measurements \citep{Zou2024}, observed i-band apparent magnitudes span $\sim$19-24\,mag, comparable to smaller samples of EELGs discovered in deep cosmological surveys at similar redshifts \citep[e.g.][]{Amorin2014,Amorin2015,Calabro2017, Ly2015,delmoral2024}. 
Most galaxies in the sample have small angular sizes, with a median effective radius of 0.4$''$ (0.25$''$ and 1.9$''$ for percentiles p16 and p95, respectively), according to Sérsic models \citep{Zou2024}. This ensures that the DESI fiber spectra (1.5$''$ size) have minimal flux losses. Using the EELGs redshifts, these radii translate to median physical sizes of 1.99 kpc (0.68 kpc and 3.66 kpc, for percentiles p16 and p84, respectively). 

In Table~\ref{tab:eelg_catalog_preview}, we present the catalog of EELGs, which we will discuss in subsequent sections after presenting the derivation of emission-line measurements and physical properties.

\section{Spectrophotometric measurements} \label{sec:methods}

To characterize the EELG population identified in DESI, we constructed a homogeneous catalog of emission-line measurements and global galaxy properties derived from spectrophotometric SED fitting. They are presented in Tables~\ref{tab:eelg_catalog_preview} and ~\ref{tab:eelg_catalog_physprop_preview}. This approach allows us to combine robust nebular diagnostics with physically motivated constraints on stellar populations and star formation histories. The main steps of the analysis are summarized below, while additional methodological details are provided in the Appendices (Sect.{app:cat}).

\subsection{Spectrophometric SED fitting}

We derived stellar population properties for the  k-MENDEL EELG sample using the SED-fitting code \textsc{CIGALE} (v2025.1; \citealt{Burgarella2005,Noll2009,Boquien2019}). Since broadband photometry is not uniformly available for all DESI targets, we constructed low-resolution spectral energy distributions by convolving each rest-frame DESI spectrum with a set of 50 box-shaped medium-band filters of width 124\,\AA, following a strategy similar to that adopted by \citet{Zou2024} and inspired by narrow-band analyses in J-PAS EELG samples \citep[e.g.][]{Iglesias2022,Breda2024,Gimenez-Alcazar2025}. This procedure preserves the global continuum shape while reducing sensitivity to noise and narrow spectral features.

SED fitting was performed at fixed spectroscopic redshift using a grid of composite stellar population models that include both an evolved underlying component and a young burst, following \citet{Lumbreras-Calle2022}. Stellar emission was modeled using the Charlot \& Bruzual (2019) libraries with a \citet{Chabrier2003} IMF and metallicities ranging from 5\% solar to solar. The old population accounts for the diffuse stellar component commonly observed in compact star-forming dwarfs and Green Pea galaxies \citep{Papaderos1996b,Amorin2009,Amorin2012}, with ages between 0.05 and 5\,Gyr, while the young burst component represents ongoing star formation episodes with ages between 1 and 12\,Myr. The burst mass fraction was allowed to vary between 2.5\% and 20\%, consistent with previous studies of EELG star formation histories \citep[e.g.][]{Amorin2012,Fernandez2022}. Exponentially declining star formation histories ($\tau$-models) with characteristic e-folding timescales of 1 and 10\,Myr were adopted for the young and old components, respectively.

Nebular continuum and line emission were included using the photoionization models of \citet{Inoue2014}, while dust attenuation was modeled using the \citet{Calzetti2000} extinction law. We explored a wide range of ionization parameters and gas-phase metallicities, assuming a representative electron density of $n_e=100$\,cm$^{-3}$. For simplicity, the escape fraction of ionizing photons was fixed to zero, an assumption that may not hold for some EELGs but has limited impact on the derived stellar masses and global SED properties. A summary of the adopted parameter grid is provided in Table~\ref{cigale_tab}. 
      
\begin{figure}[!t]
    \centering
    \includegraphics[width=0.49\textwidth]{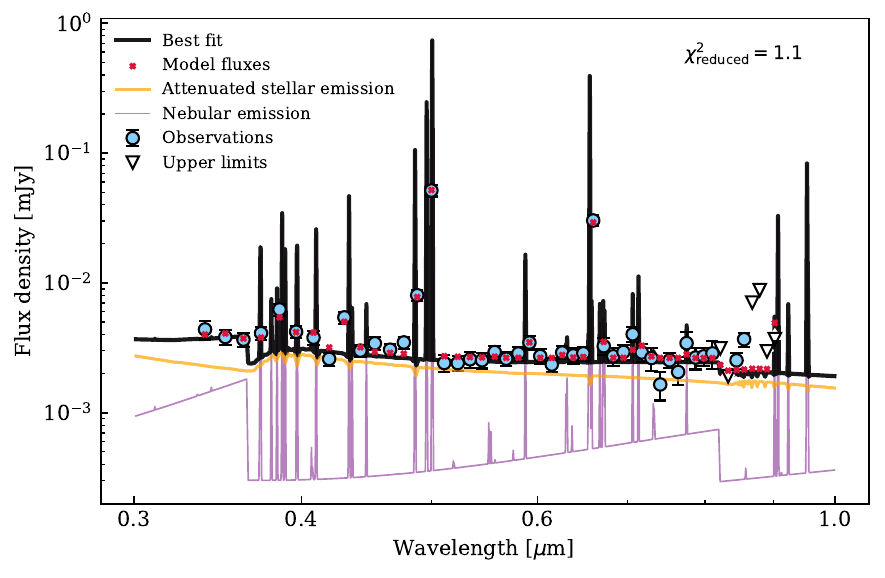}
    \caption{Example of spectrophotometric SED fitting with \textsc{CIGALE} for a representative EELG (DESI TARGETID\,39633286554487272). Colored symbols show the synthetic medium-band photometric points derived from the DESI spectrum after convolution with box filters of 124\,\AA\ width, while the solid black curve indicates the best-fit model including stellar, nebular, and dust components. The fit illustrates the ability of the method to reproduce both the continuum shape and the contribution of strong emission lines in low signal-to-noise spectra.}
    \label{fig:SED}
\end{figure}

Figure~\ref{fig:SED} shows an example of the resulting SED fit for a representative galaxy. The main derived physical quantities, including stellar mass, star formation rate averaged over the last 10\,Myr (SFR$_{10}$), stellar population ages, and dust attenuation, are included in the catalog described in Section~\ref{app:cat}. We note that the use of medium-band synthetic photometry mitigates spectrophotometric calibration uncertainties between DESI spectrograph arms (see Section~\ref{app:cat}), providing robust relative constraints on galaxy physical parameters despite the moderate signal-to-noise of individual continua.

\subsection{Emission line measurement with LiMe}

Emission lines were measured from DESI spectra using the LIne MEasuring library for large and complex spectroscopic datasets (LiMe; \citealt{LiMe2024}). This tool performs automated line detection and profile fitting based on Gaussian models and non-linear least-squares optimization using the \texttt{LMFIT} package \citep{Newville2016}. We fitted all emission lines detected above a 2$\sigma$ significance threshold, covering rest-frame wavelengths from [Mg\,{\sc ii}]$\lambda$2798 to [S\,{\sc iii}]$\lambda$9531 depending on galaxy redshift. Due to the typically low signal-to-noise ratio of the stellar continuum in EELG DESI spectra, we did not perform a global stellar continuum subtraction. Instead, LiMe fits a first-order polynomial to the local continuum around each emission line. This approach minimizes systematic biases introduced by full spectral synthesis fitting when continuum constraints are weak \citep[e.g.][]{FADO2017}.

Equivalent widths were corrected for underlying stellar absorption by applying a uniform correction of 2\,\AA\ to the \hb\ line, consistent with typical values observed in EELGs \citep[e.g.][]{Ly2015,Amorin2014}. While this simplified treatment neglects object-to-object variations, the impact on the strongest emission lines is small compared to measurement uncertainties. Moreover, diffuse ionized gas contamination and stellar absorption effects are expected to be minor in extreme emission-line samples \citep[e.g.][]{Ly2015,Sanders2017}.

The resulting emission-line flux catalog is presented in Section~\ref{app:cat}. Overall, our flux ratios show good agreement with measurements from the DESI EDR value-added catalog of \citet{Zou2024}, while our equivalent widths are systematically larger due to differences in continuum estimation methods. The EW$_{5007}$ distribution spans $\sim$100–2000\,\AA\ with a mean value of 420\,\AA, while \hb\ equivalent widths range from $\sim$20\,\AA\ to 500\,\AA\ (mean 88\,\AA), confirming the extreme nature of the  sample.

A key result of our measurements is the detection of intrinsically faint auroral lines, which enable direct electron-temperature metallicity determinations. In particular, the [\oiii]$\lambda$4363 line is detected with S/N$>$2 (S/N$>$3) in 3499 (1708) galaxies, corresponding to $\sim$21.5\% (10.5\%) of the full sample. This large statistical sample of $T_e$-based measurements represents a significant improvement over previous studies based on smaller spectroscopic datasets and forms the basis for the metallicity analysis presented in Section~\ref{sec:diagnostics}.

\begin{figure*}[!t]
    \centering
\includegraphics[width=0.49\textwidth]{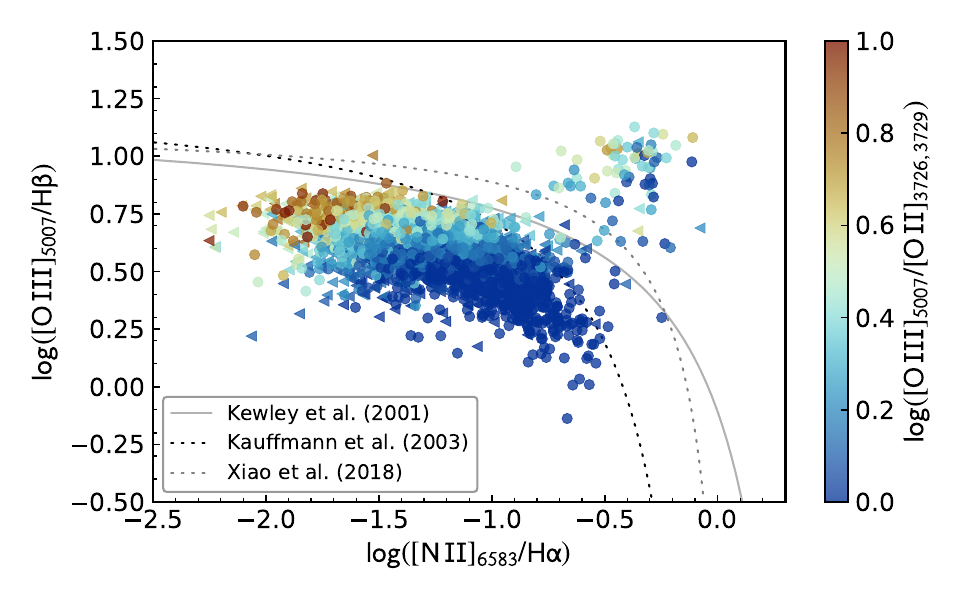}
\includegraphics[width=0.49\textwidth]{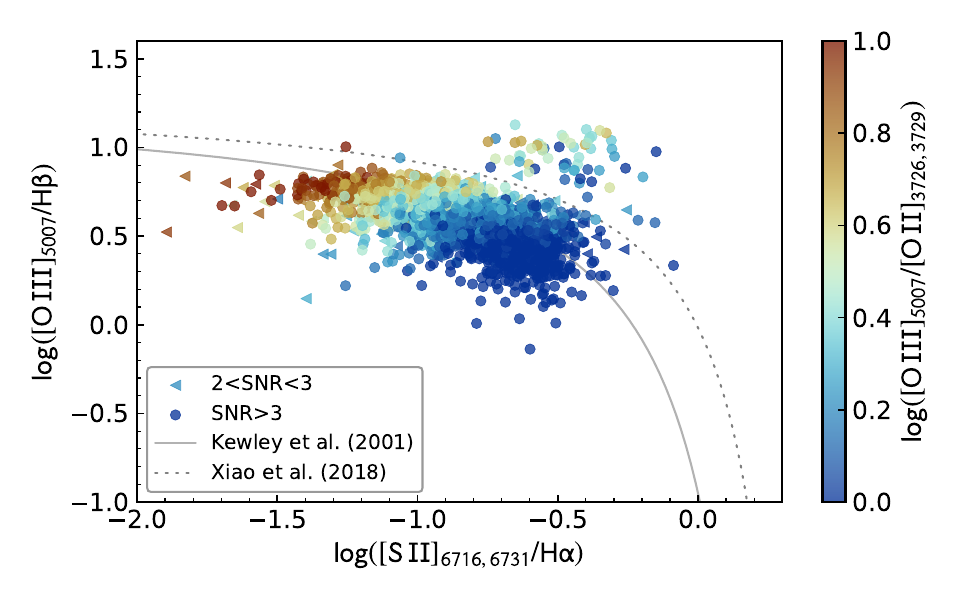} \\
\includegraphics[width=0.49\textwidth]{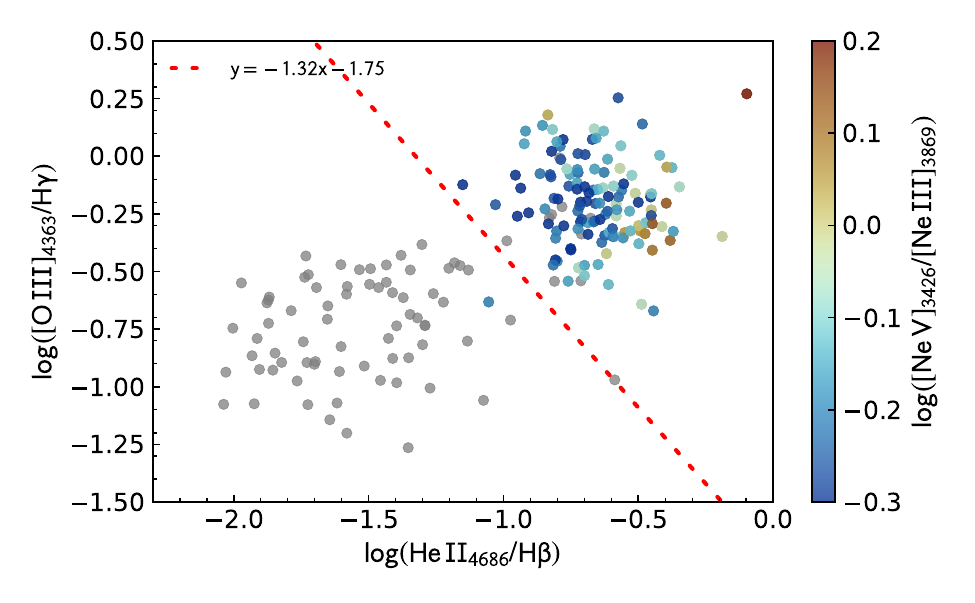}
\includegraphics[width=0.49\textwidth]{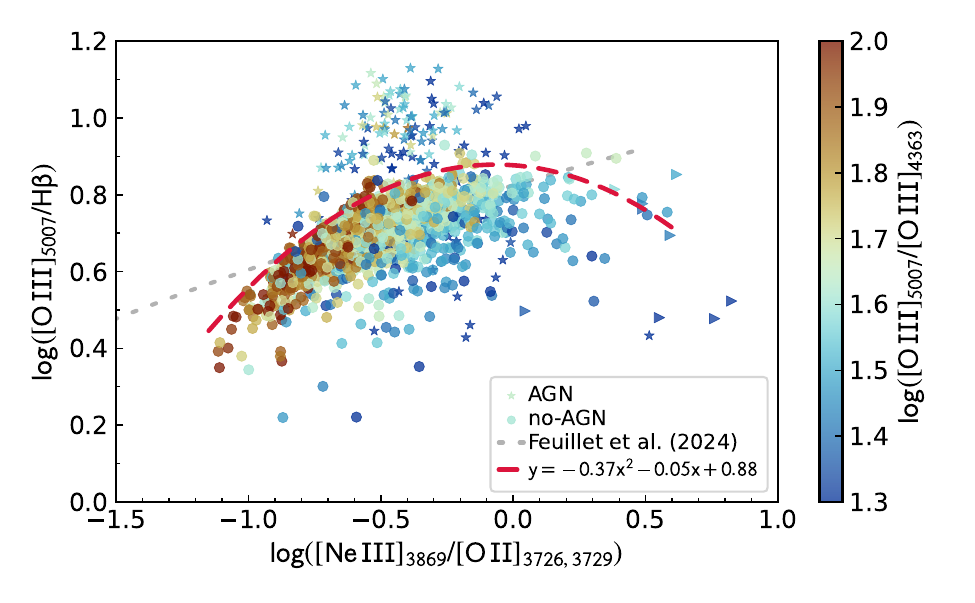}
    \caption{Ionization diagnostic diagrams for the DESI EELG sample. \textit{Top panels:} Classical BPT diagrams for the subsample at $z<0.45$, for which [N\,{\sc ii}]$\lambda6583$ (left) and [S\,{\sc ii}]$\lambda\lambda6716,6731$ (right) are accessible. Solid, dashed, and dotted curves show the demarcation relations from \citet{Kewley2001}, \citet{Kauffmann2003}, and \citet{Xiao2018}, respectively. Symbols are colour-coded according to the $O_{32}$ ratio, tracing variations in ionization parameter. \textit{Bottom left:} O3H$\gamma$ versus He\,{\sc ii}$\lambda4686$/H$\beta$ diagram highlighting the location of high-ionization systems. Grey points indicate EELGs with S/N$<2$ in He\,{\sc ii}. The red dashed line shows the empirical demarcation adopted in this work to separate extreme ionization sources (see text). \textit{Bottom right:} [O\,{\sc iii}]$\lambda5007$/H$\beta$ versus Ne3O2 diagnostic diagram. Green symbols indicate AGN candidates identified from composite ionization criteria, while blue symbols correspond to star-forming EELGs. The dashed red curve shows the empirical relation proposed by \citet{Feuillet2024}. In all panels, colour bars indicate line-ratio diagnostics sensitive to ionization conditions.
}
    \label{fig:BPT}
\end{figure*}

\subsection{Interstellar extinction correction from Balmer lines}\label{sec:extinction}

Emission-line fluxes were corrected for interstellar extinction using the reddening coefficient $c(H\beta)$ and adopting the extinction curve of \citet{Cardelli1989}. Intrinsic line intensities were derived following
\begin{equation}
\log \left(\frac{I_{\lambda}}{I_{H\beta}}\right)
=
\log \left(\frac{F_{\lambda}}{F_{H\beta}}\right)
+
c(H\beta)\,f(\lambda),
\end{equation}

The reddening coefficient was estimated from the brightest available Balmer ratios, primarily H$\alpha$/H$\beta$ and H$\gamma$/H$\beta$, depending on redshift coverage and requiring S/N$\geq$3 in the relevant lines. Observed ratios were compared with theoretical case B recombination values (H$\alpha$/H$\beta$\,=\,2.78 and H$\gamma$/H$\beta$\,=\,0.468), assuming representative physical conditions of the ionized gas ($T_e = 1.5\times10^{4}$\,K and $n_e = 100$\,cm$^{-3}$).

For starburst-dominated EELGs (see Section~\ref{sec:diagnostics}), we find a median reddening coefficient $c(H\beta)_{\rm med}=0.18$ (corresponding to $E(B-V)=0.12$\,mag), while galaxies with AGN-like ionization signatures show higher median values of $c(H\beta)_{\rm med}=0.31$ ($E(B-V)=0.21$\,mag). Individual extinction values are provided in the emission-line catalog.

A non-negligible fraction of galaxies exhibits Balmer ratios slightly below the theoretical case B values. While measurement uncertainties can partly explain these deviations, we adopt $c(H\beta)=0$ for such objects. These low ratios may also reflect departures from standard assumptions, including variations in electron temperature and density, deviations from pure case B recombination, or Balmer self-absorption effects. Similar behaviour has been reported in recent studies of metal-poor, compact star-forming galaxies \citep[e.g.][]{Scarlata2024,Yanagisawa2024,McClaymont2025}. A dedicated analysis of these systems will be presented in future work.

\section{Characterization of the k-MENDEL EELG sample} \label{sec:results}

In this section we use the derived emission-line measurements and spectrophotometric properties to characterize the EELG population identified in DESI. We focus on global trends revealed by ionization diagnostics, composite spectra, scaling relations, and gas-phase metallicities. For the following analysis we restrict the sample to the redshift range $0.01<z<0.85$ in order to avoid possible systematic uncertainties in line ratios caused by large telluric residuals at the red end of the spectra. 

\subsection{Emission-line diagnostics and high ionization lines: The AGNs contribution to the EELG sample}\label{sec:diagnostics}

We first investigate the dominant ionizing sources in the EELG sample using a set of classical and high-ionization emission-line diagnostics. In Fig.~\ref{fig:BPT} we present standard BPT diagrams involving [\nii], [\sii], and [\oiii] line ratios \citep{Baldwin1981,Veilleux1987}. These diagnostics can be applied to galaxies at $z\lesssim0.5$, where the required emission lines remain within the DESI spectral range.

To extend the analysis to higher redshift and lower metallicity regimes, we also use diagnostics based on [\oii] and [\neiii] ratios \citep[e.g.][]{Zeimann2015,Backhaus2022,Backhaus2025}, as well as the presence of high-ionization lines such as \heii\,$\lambda4686$ and [\nev]\,$\lambda3426$. These transitions trace hard radiation fields that may arise from very massive stars, shocks, or accreting black holes \citep[e.g.][]{Kehrig2015,Cleri2023,Mingozzi2024}. 

From these diagnostics we identify 942 galaxies with line ratios inconsistent with pure stellar photoionization models. These objects likely host additional ionizing sources, including shocks and narrow- or broad-line AGNs. They represent $4.8\%$ of the EELG sample at $0.01<z<0.85$, a fraction somewhat lower than the $\sim10\%$ reported in smaller EELG samples \citep[e.g.][]{Amorin2015,Calabro2017}. This difference is likely driven by selection effects, as our ASK-based identification preferentially isolates spectra with extreme equivalent widths, whereas many AGNs populate intermediate-EW spectral classes (particularly ASK10–ASK12) that are not included in our EELG definition.

At higher redshifts, larger AGN fractions have been reported in samples of extreme line emitters \citep[e.g.][]{Davis2026}. However, differences in diagnostics, EW selection thresholds, and stellar-mass ranges complicate direct comparisons. Recent work based on DESI data has shown that AGN fractions in low-mass galaxies vary strongly with stellar mass and selection criteria \citep{Pucha2025}, further highlighting the role of sample definition.

Because the primary goal of this work is to characterize compact star-forming EELGs, we exclude AGN candidates from the subsequent analysis. Their properties will be explored in detail in a forthcoming study (Fernández-Ontiveros et al., in prep.).

\subsection{Composite spectra of DESI EELGs}

\begin{figure*}[!t]
    \centering
    \includegraphics[width=0.99\textwidth]{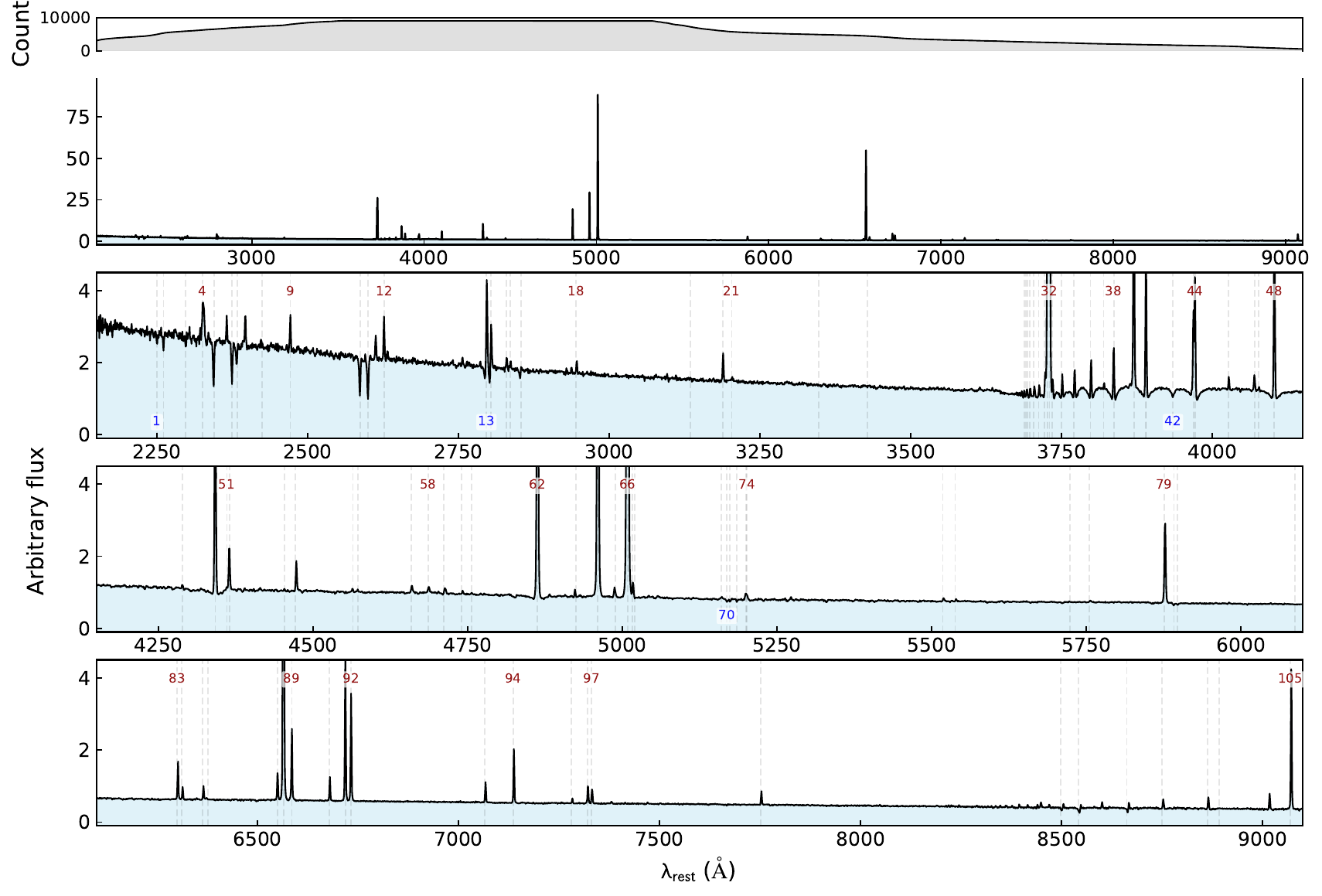}
    \caption{Median rest-frame composite spectrum of the k-MENDEL EELG sample after excluding sources consistent with AGN activity based on emission-line diagnostics (Sect.~\ref{sec:diagnostics}). The spectrum is shown in several wavelength intervals to highlight both nebular emission lines and stellar continuum features. Grey dashed vertical lines mark the spectral windows used for emission-line measurements, while red labels identify the main emission features and blue labels indicate prominent stellar absorption lines, following the notation described in Table~\ref{tab:composite}. The light-blue shaded region illustrates the underlying stellar continuum level. The upper panel shows the number of galaxies contributing to the stack as a function of rest-frame wavelength, reflecting the redshift-dependent spectral coverage of the DESI sample. The high signal-to-noise ratio of the stack allows the detection of weak stellar absorption features, revealing the presence of evolved stellar populations underlying the ongoing starburst activity.
    }
    \label{fig:median-stack}
\end{figure*}

To investigate spectral features that are too faint to be detected in individual spectra, we constructed median stack spectra of the EELG sample. Details of the stacking procedure are provided in Section~\ref{app:composite}. Figure~\ref{fig:median-stack} shows the composite spectrum for the subsample of star-forming EELGs, while the corresponding stack for AGN candidates is presented in Fig.~\ref{fig:agn-stack}.

The composite spectra reveal a wealth of nebular emission lines spanning a wide ionization range, including faint auroral transitions that are rarely measurable in individual objects. Nebular \heii\ emission is commonly detected, supporting the presence of hard radiation fields in a significant fraction of EELGs. In contrast, very high-ionization lines such as [\nev] are not detected in the star-forming composite (S/N$<2$), confirming that such features are rare in purely stellar-powered systems \citep[e.g.][]{Izotov2021_nev,Cleri2023_clear,Mingozzi2025,delValle-Espinosa2025_cgcg}. These lines are, however, clearly present in the AGN composite, along with other very high-ionization lines, e.g. [Fe\,{\sc vii}].

The high signal-to-noise of the stacked spectra also allows the identification of several faint UV absorption and emission features associated with the interstellar and circumgalactic media. In particular, we detect Fe\,{\sc ii} absorption lines at 2344, 2374, 2382, 2586, and 2600\,\AA, together with fine-structure Fe\,{\sc ii*} emission. Mg\,{\sc ii} emission is also present, tracing enriched gas at $T\sim10^{4}$\,K in the ISM and CGM \citep[e.g.][]{Erb2012,Zhu2015}. These features are typically enhanced in compact, high-sSFR galaxies with low dust attenuation and are valuable diagnostics of gas flows \citep{Erb2012,Kornei2013}. The large DESI EELG sample therefore provides an opportunity to statistically explore outflow and feedback signatures and compare them with those observed in higher-redshift galaxies \citep[e.g.][]{Kehoe2025}.

Finally, the composite spectra offer insight into the star formation histories of EELGs. In addition to narrow nebular emission lines, we detect Balmer absorption wings as well as stellar features such as the Ca\,{\sc ii} triplet and Mg\,{\sc i}. These signatures indicate that most EELGs host an evolved stellar component with ages of several Gyr, consistent with previous studies of blue compact dwarf and green pea galaxies \citep{Papaderos1996b,Amorin2009,Amorin2012,Fernandez2022}. While a detailed reconstruction of star formation histories is beyond the scope of this paper, the statistical power of DESI highlights the potential of stacking analyses to address the star formation history of EELGs through detailed spectral synthesis fitting. 

\begin{figure*}[!t]
    \centering
\includegraphics[width=0.98\textwidth]{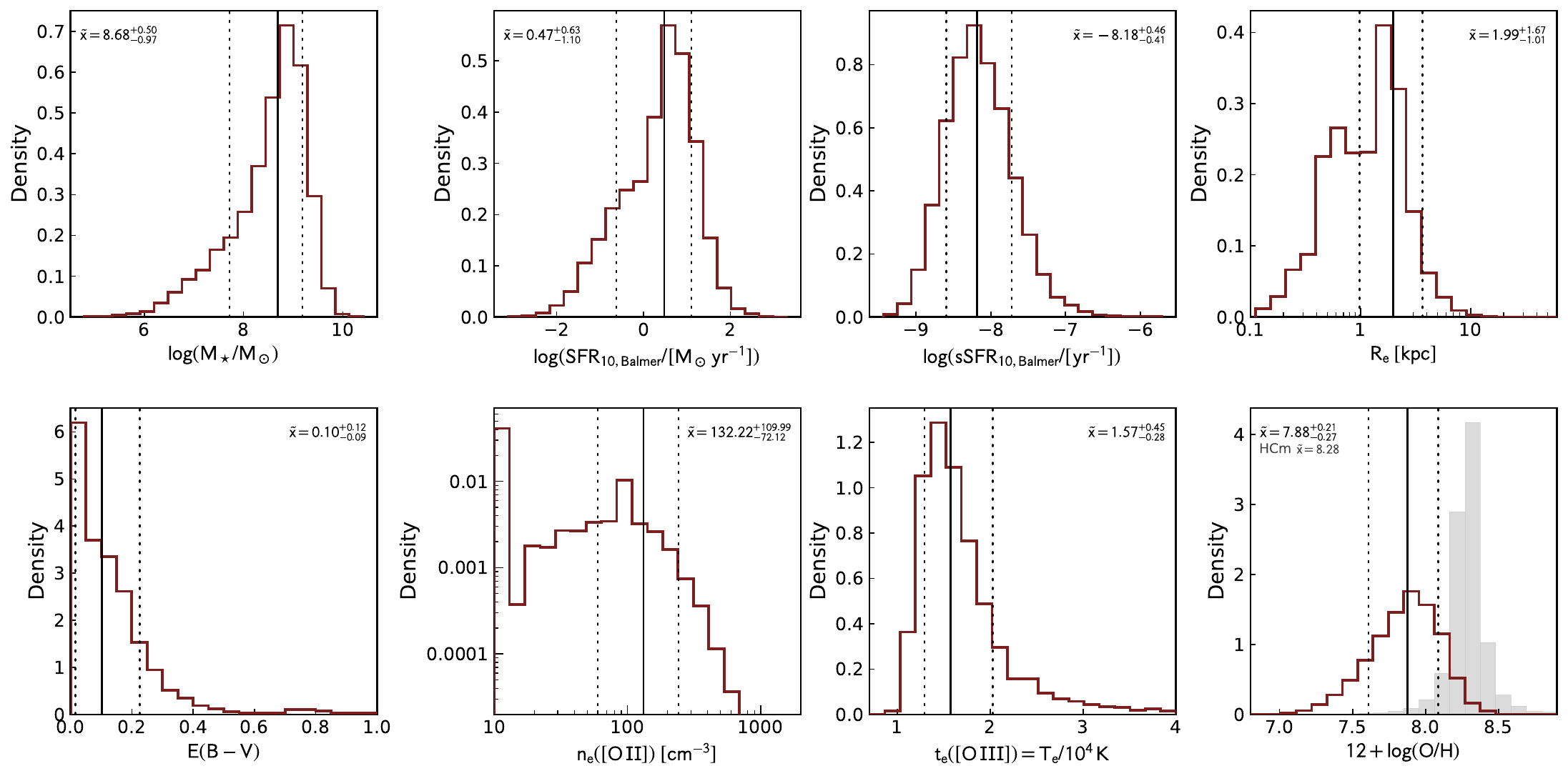}
    \caption{Distributions of physical and nebular properties for the k-MENDEL EELG sample. From left to right and top to bottom we show stellar mass, Balmer-based SFR, specific SFR, effective radius, nebular color excess $E(B-V)$, electron density $n_e$ from [O\,{\sc ii}], electron temperature $t_e$ from [O\,{\sc iii}], and oxygen abundance derived from the direct $T_e$ method. Solid vertical lines indicate the median values of each distribution, while dashed lines mark the 16th and 84th percentiles. Annotated labels provide the corresponding percentile ranges. In the metallicity panel (bottom right), the shaded grey histogram represents HCm-based metallicities for EELGs without $T_e$ measurements.
    }
    \label{fig:histogram}
\end{figure*}

\subsection{Star formation rate and stellar mass}\label{sec:SFR}

We derive instantaneous star formation rates (SFRs) from extinction-corrected Balmer-line luminosities following \citet{Shapley2023}, assuming subsolar metallicity and a \citet{Chabrier2003} IMF. For galaxies at low redshift with S/N(\ha)$>3$ we use the \ha\ luminosity, while for higher-redshift objects we adopt the \hb\ luminosity. Aperture corrections are applied following \citet{Zhou2023}. Overall, Balmer-based SFRs are in good agreement with the SFR$_{10}$ values obtained from SED fitting. These measurements are included in the catalog described in Section~\ref{app:cat}.

\begin{figure*}[!t]
    \centering
\includegraphics[width=0.49\textwidth]{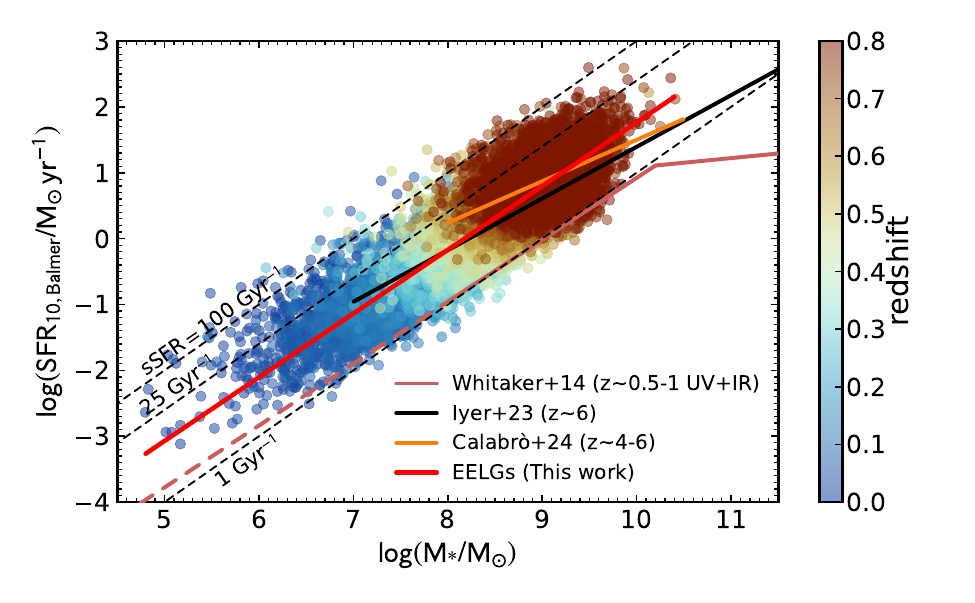}
\includegraphics[width=0.49\textwidth]{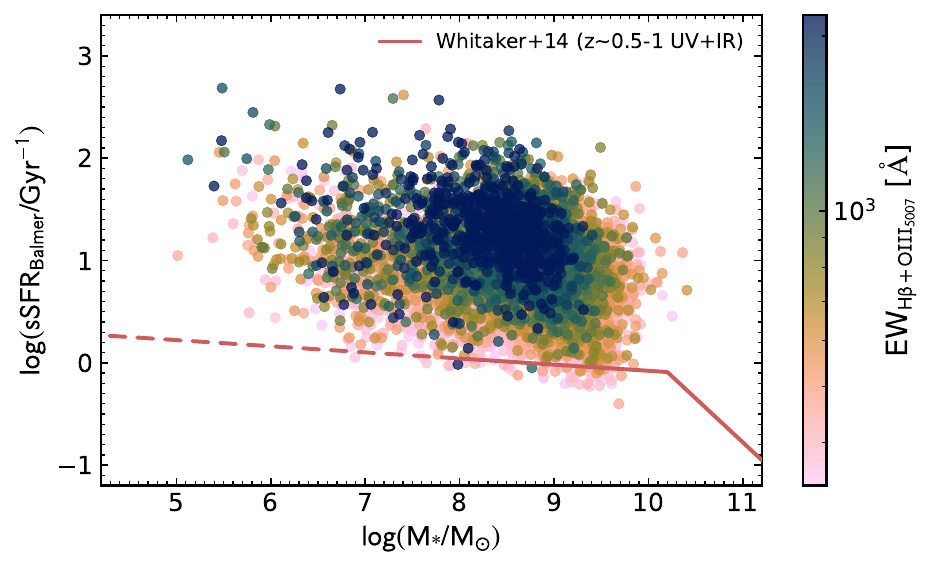}
    \caption{Star-formation rate and specific SFR vs. stellar mass relation for the k-MENDEL EELG sample. Left: galaxies colour-coded by redshift. Right: galaxies colour-coded by the combined equivalent width EW(H$\beta$+[O\,{\sc iii}]). Star-formation rates are derived from Balmer-line luminosities corrected for extinction and aperture effects. Black dashed lines indicate loci of constant specific star-formation rate (sSFR). For comparison, we show representative star-forming main-sequence relations from the literature: the local UV+IR relation from \citet{Whitaker2014} (brown curve), the $z\sim6$ relation from \citet{Iyer2018} (black line), and the $z\sim4$-6 relation from \citet{Calabro2024} (orange line). The red line shows the best-fit sequence defined by EELGs in this work. EELGs populate the upper envelope of the SFR-$M_{\star}$ plane, with the highest-EW systems reaching the largest sSFR values, consistent with increasingly young and intense starburst episodes.}
    \label{fig:SFR-M*}
\end{figure*}

Figure~\ref{fig:SFR-M*} shows the stellar mass-SFR relation for the EELG sample. EELGs populate a relatively narrow sequence systematically located above the star-forming main sequence at similar redshift \citep{Whitaker2014}. This corresponds to high specific star formation rates in the range $\sim10$-$100$\,Gyr$^{-1}$, consistent with previous studies of EELGs at comparable redshifts \citep{vanderWel2011,Amorin2015}. These values support the interpretation that most systems are undergoing short-lived, intense starburst episodes rather than sustained star formation.

A clear redshift-dependent trend is visible in Fig.~\ref{fig:SFR-M*}, reflecting the Malmquist bias of the DESI survey, which limits the detection of lower-mass and lower-SFR EELGs at higher redshifts. In addition, we find that galaxies with larger emission-line equivalent widths tend to occupy the upper envelope of the mass-SFR relation. Despite the relatively large uncertainties in EW measurements due to faint continua, this trend indicates that the youngest and most intense bursts of star formation correspond to the highest specific SFR values. 

\subsection{Gas-phase metallicity}\label{sec:abundances}

Gas-phase metallicity provides a key constraint on the evolutionary stage of EELGs and their role as analogs of high-redshift star-forming systems. In the following, we derive oxygen abundances using both the direct $T_e$ method, when auroral lines are available, and strong-line diagnostics calibrated for extreme emission-line conditions. This dual approach allows us to characterize the metallicity distribution of the full sample while preserving a robust anchor to direct abundance determinations.

\subsubsection{Direct $T_e$ metallicities}\label{sec:oxygen}

We derive gas-phase oxygen abundances using the direct $T_e$ method, based on electron temperature and density measurements from reddening-corrected emission-line ratios. Calculations were performed with \textsc{PyNeb} \citep{Luridiana2015}. Electron density ($n_e$) was obtained from the [\oii]$\lambda\lambda$3727,3729 ratio ($R_{\rm O2}$), which is accessible over the full redshift range of the sample, assuming an initial temperature of $10^{4}$\,K. The electron temperature of the high-ionization zone was derived from the $R_{\rm O3}=$([\oiii]$_{4959,5007}$/[\oiii]$_{4363}$) ratio. The temperature of the low-ionization zone was then estimated following the formulation of \citet{Hagele2008}, which accounts for its dependence on electron density.

Ionic abundances were computed using collisional strengths from \citet{Kisielius2009} for O$^{+}$ and \citet{Storey2014} for O$^{2+}$, and total oxygen abundances were obtained by summing both ionic contributions. Uncertainties in $n_e$, $T_e$, and abundances were estimated through Monte Carlo propagation of line-intensity errors using 1000 realizations.

Out of the k-MENDEL EELG sample, 3039 star-forming galaxies satisfy the requirement of having the relevant diagnostic ratios measured with S/N$>2$. For a small subset lacking reliable measurements of the [\oii] doublet, we adopt $n_e = 100$\,cm$^{-3}$ and estimate metallicities using the empirical relation between $T_e$(\oiii) and oxygen abundance calibrated for EELGs by \citet{Amorin2015}, which shows a dispersion of $\sim0.15$ dex and is valid in the range $t_e$(\oiii)$\sim1$–2.5 \citep[see also][]{EPM2021}.

Figure~\ref{fig:histogram} shows the distributions of $n_e$ and $T_e$ for galaxies with detected [\oiii]$\lambda4363$. We obtain median (16th–84th percentile) values of $n_e = 132^{+72}_{-110}$\,cm$^{-3}$ and $T_e = 1.56^{+0.49}_{-0.49}\times10^{4}$\,K. Direct-method metallicities span $12+\log(\mathrm{O/H})\sim6.9$–8.5, with a median value of 7.85 ($\sim$\,15\% solar), consistent with previous studies of EELGs based on smaller and typically brighter samples \citep[e.g.][]{Amorin2010,EPM2021}.


\subsubsection{Strong-line metallicities}\label{sec:metallicity}

For the majority of EELGs, auroral lines are not detected due to limited signal-to-noise. In these cases, we derive gas-phase metallicities using the HII-Chi-Mistry code \citep[HCm;][]{Perez-Montero2014}. This approach relies on photoionization models to simultaneously constrain O/H, N/O, and the ionization parameter by fitting all available strong emission-line ratios in a self-consistent framework tied to the direct $T_e$ scale.

We adopt the version of HCm specifically optimized for EELGs \citep{EPM2021}, which is based on {\sc cloudy} v17 models \citep{Ferland2017} and BPASS ionizing spectra. Input line intensities, normalized to \hb, include [\oii]$\lambda\lambda3727,3729$, [\neiii]$\lambda3869$, [\oiii]$\lambda4363$, [\oiii]$\lambda\lambda4959,5007$, [\nii]$\lambda6584$, and [\sii]$\lambda\lambda6717,6731$, requiring S/N$>3$. When [\oiii]$\lambda4363$ is not detected, HCm uses alternative model grids depending on the availability of [\nii] and [\sii] lines \citep[][for details]{EPM2021}. Typical uncertainties in the derived metallicities are $\sim0.2$ dex, comparable to other widely used strong-line calibrations \citep[e.g.][]{Curti2020,Nakajima2023}.

Applying HCm to the full k-MENDEL EELG sample yields metallicities in the range $12+\log(\mathrm{O/H})\sim7.0$–8.6, with a median value of 8.18 ($\sim$\,30\% solar), consistent with previous studies of EELGs in SDSS \citep[e.g.][]{Amorin2010,EPM2021}. For galaxies without auroral-line detections, the median metallicity increases to $12+\log(\mathrm{O/H})\sim8.3$, as shown in Fig.~\ref{fig:histogram}. A more detailed comparison between strong-line diagnostics and direct-method abundances in EELGs will be presented in future work.

\section{Discussion: EELGs as powerful laboratories for early chemical evolution and cosmic reionization}\label{sec:discussion}

Extreme emission-line galaxies are increasingly interpreted as short-lived but recurrent evolutionary phases in the growth of low-mass galaxies, characterized by intense star formation, high ionization parameters, and low gas-phase metallicities.
Systems identified at different redshifts under various observational labels,  including blueberry galaxies, green peas, and luminous compact galaxies, share the defining property of large nebular equivalent widths and therefore represent manifestations of the same physical phenomenon observed under different selection conditions \citep[e.g.][]{Cardamone2009,Amorin2010,Izotov2011,Yang2017_bb,EPM2021}.

Recent \textit{JWST} observations have revealed that galaxies with similarly extreme nebular properties are common at $z>4$, suggesting that the EELG phase may represent a dominant mode of star formation in the early Universe \citep[e.g.][]{Schaerer2022,Rhoads2023,Llerena2024}. 
In this context, large statistical samples of local analogs provide a unique opportunity to investigate the physical processes regulating star formation, chemical enrichment, and ionizing-photon production under conditions similar to those prevailing during cosmic reionization.

The k-MENDEL EELG sample extends previous spectroscopic studies to significantly larger numbers and wider dynamic ranges in stellar mass, star formation rate, and metallicity. 
This enables a statistically robust exploration of scaling relations and ionization properties, helping to place extreme starbursts within the broader framework of galaxy evolution. 
In particular, the combination of large sample size and direct-method metallicity measurements provides new constraints on the nature of departures from equilibrium growth in low-mass galaxies.

\subsection{The mass-metallicity-SFR relation of EELGs with direct metallicities}\label{sec:MZR}

\begin{figure*}[!t]
    \centering
\includegraphics[width=0.99\textwidth]{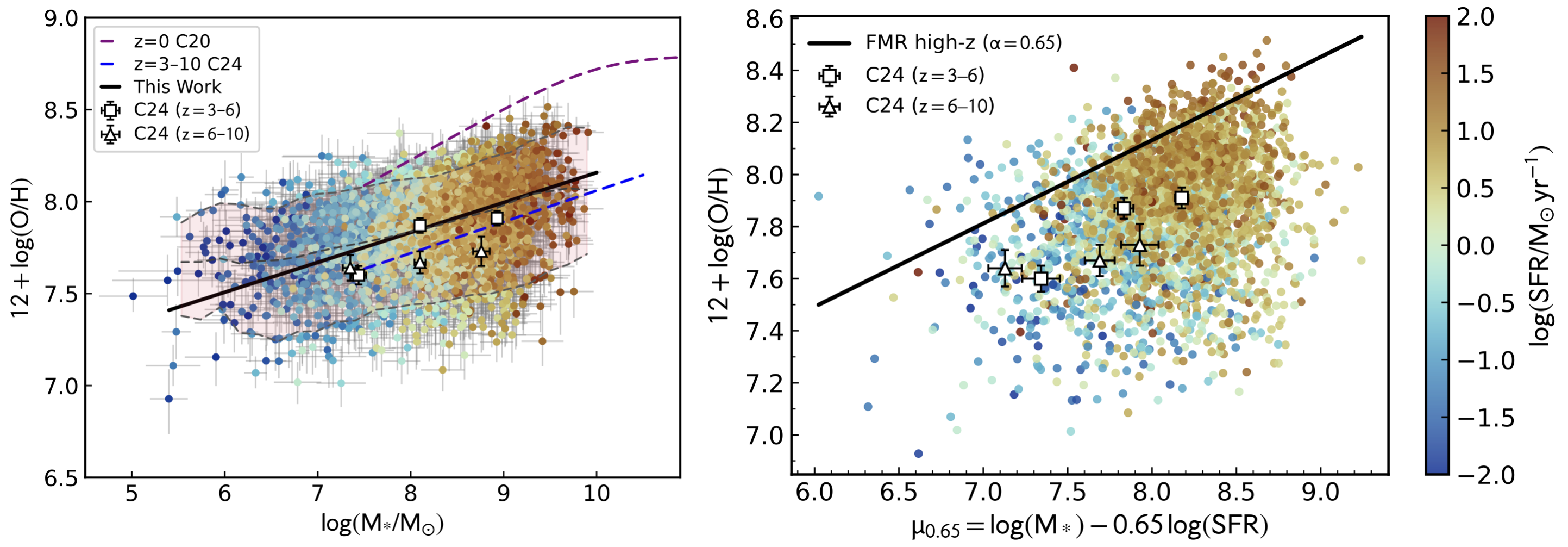}
    \caption{
Mass–metallicity relation (left) and FMR projection (right) for EELGs with direct metallicity measurements based on the detection of the [\oiii]$_{4363}$ auroral line. 
Individual galaxies are shown as colour-coded circles according to their instantaneous SFR derived from aperture-corrected fibre spectra. In the top panel, the black solid line shows the best-fit MZR relation obtained in this work. Grey dashed curves indicate the median binned relation of the EELG sample together with the 5th and 95th percentile scatter envelope. For comparison, we show the local Te-based MZR for normal star-forming galaxies from \citet{Curti2020} (magenta dashed curves) and the MZR fit derived for \textit{JWST} galaxies at $z\sim3$–10 from \citet{Curti2024_mzr} (black dashed line). 
The left panel shows the projection of the same galaxies onto the FMR parameter $\mu_{0.65}=\log(M_{\star})-0.65\log(\mathrm{SFR})$. The solid black line indicates the high-EW FMR parametrization from \citet{Curti2024_mzr}. Square and triangular symbols show the median binned measurements reported in that work for galaxies at $z\sim3$-6 and $z\sim6$-10, respectively. This projection illustrates that the large metallicity scatter of EELGs is not significantly reduced when accounting for SFR.
}
    \label{fig:MZR}
\end{figure*}

EELGs are commonly interpreted as rapidly evolving systems undergoing short-lived starbursts episodes, characterized by very high sSFRs, extreme EWs, and low gas-phase metallicities. They typically lie on the upper envelope of the SFR-$M_{\star}$ plane, well above the star-forming main sequence, and are offset toward lower metallicities relative to the average galaxy population  \citep[e.g.][]{Amorin2010,Calabro2017,Boyett2024,Llerena2024,Katz2025}. 
These properties suggest that EELGs trace phases of accelerated growth rather than steady evolution.

Figure~\ref{fig:MZR} shows the mass-metallicity relation (MZR) defined by the subsample of EELGs with direct metallicities. Over the stellar mass range probed by our sample, EELGs follow a shallow relation, 
\begin{equation}
    12+\log(O/H) =  (0.16 \pm 0.01) \times (\log(M_{\star}) - 8) + (7.83 \pm 0.01) 
\end{equation}
which is systematically offset by $\sim$0.3-0.5 dex toward lower metallicities compared to local star-forming galaxies \citep[e.g.][]{Andrews2013,Curti2020}. 
The slope and normalization of this relation are consistent with previous studies of auroral-line selected low-mass galaxies at similar reedshifts \citep[e.g.][]{Berg2012,Berg2022,Calabro2017}. 

For comparison, Fig.~\ref{fig:MZR} also includes the median relations derived for high-redshift galaxies observed with \textit{JWST} by \citet{Curti2024_mzr}. 
Consistent trends are also reported by other direct-metallicity studies at high redshift based on \textit{JWST} observations \citep[e.g.][Giménez-Alcázar, submitted]{Nakajima2023,Chemerynska2024,Morishita2024}. 
Remarkably, the locus defined by local EELGs overlaps with the binned measurements reported for galaxies at $z\sim3$-10, suggesting that low-redshift extreme emitters share similar chemical scaling relations with rapidly assembling galaxies in the early Universe. 

Despite this global agreement, the intrinsic scatter of the EELG MZR remains large ($\sim$1 dex), significantly exceeding that observed in normal star-forming galaxies \citep[e.g.][]{Andrews2013,Curti2020}. 
Such large dispersion is naturally expected in systems where stochastic gas accretion, rapidly varying star-formation efficiencies, and strong feedback operate on short timescales.

\begin{figure*}[!t]
    \centering
\includegraphics[width=0.9\textwidth]{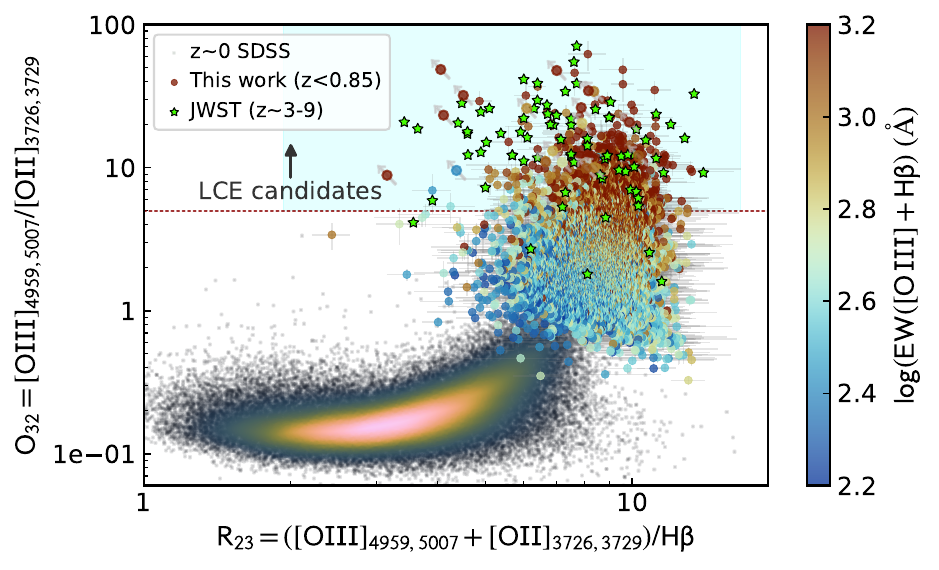}
    \caption{
$O_{32}$ versus $R_{23}$ diagram for the DESI EELG sample. Colored circles show individual EELGs at $z<0.85$, color-coded by $\log$ EW(H$\beta$+[O\,{\sc iii}]). 
The grey density map represents normal star-forming galaxies at $z<0.1$ from SDSS DR16 \citep{EPM2021}. 
Green star symbols indicate metal-poor galaxies at $z\sim3$-9 observed with \textit{JWST}. 
The horizontal dashed line marks the approximate regime of candidate LyC emitters ($O_{32}\gtrsim5$). 
DESI EELGs populate a high-ionization sequence clearly offset from the bulk of local star-forming galaxies and overlapping with the locus of reionization galaxies.}
    \label{fig:O32-R23}
\end{figure*}

A natural framework to interpret metallicity variations is the fundamental metallicity relation (FMR), which links stellar mass, metallicity, and star-formation rate in simple stationary or “bathtub” models of galaxy evolution \citep[e.g.][]{Mannucci2010,Lilly2013}. In the FMR framework, an anticorrelation between metallicity and SFR is observed for galaxies of a given stellar mass. 
Analytical models and numerical simulations explain the FMR as an increased SFR associated with recent time-variable gas inflows that dilute metallicity while sustaining star formation, leading to a reduction of the MZR scatter \citep{SanchezAlmeida2014_araa}. 

In Figure~\ref{fig:MZR}(right), we show the projection of the EELG sample along the $\mu_{0.65}$ parameterization of the FMR for high-EW local galaxies in \citet{Curti2020}. While a trend is shown, EELGs remain systematically offset by $\sim$0.3 dex from the FMR followed by galaxies out to $z\sim$\,2.5 \citet{Curti2024_mzr}, and the intrinsic scatter remains large, reaching up to $\sim$\,1 dex over the full range of stellar mass and SFR. 
This indicates that the SFR alone is not the primary driver of metallicity variations in these systems. The median binned measurements reported for galaxies at $z\sim3$-10 are broadly consistent with the locus defined by our sample, reinforcing the similarity between local EELGs and rapidly assembling galaxies in the early Universe.
 
Overall, these results support a scenario in which EELGs experience strong departures from the stationary or steady state evolution predicted by models and simulations predicting the FMR, suggesting that their baryon cycle is dominated by rapid, episodic processes such as stochastic and massive inflows of metal-poor gas fueling short-lived starbursts and driving intense feedback. 
Such processes can produce coherent offsets from local scaling relations and sustain low metallicities despite high star-formation efficiencies, as previously suggested for green peas and other EELGs at low- and intermediate-redshift \citep[e.g.][]{Amorin2010,Amorin2014,Amorin2017,Calabro2017}. 
The similarity with recent \textit{JWST} results \citep[e.g.][]{Nakajima2023,Curti2024_mzr,Morishita2024,Sarkar2025} indicates that these extreme phases in which galaxies are out from stationary-state predicted by models predicting the FMR, may be common in low-mass galaxies across cosmic time. 

In this sense, our results show that our k-MENDEL EELGs may represent local realizations of the dominant highly efficient star-formation mode at early cosmic epochs, in which star formation proceeds faster than the build-up of metals, producing the large offsets observed in galaxy scaling relations. The reasons why these low-redshift, low-mass galaxies are still affected by a time-variable massive inflow rate of gas triggering efficient star formation and leading to lower metallicities otherwise predicted by models needs further investigation. 
Overall, the similarity between the scaling relations of local EELGs and those observed at high redshift suggests that the physical mechanisms governing early galaxy assembly may still operate in low-mass galaxies at later cosmic times. 
This further indicates that extreme emission-line selection efficiently isolates galaxies in highly non-equilibrium evolutionary phases. 
While bathub models are primarily designed to reproduce average trends rather than the full distribution of galaxy properties, reproducing EELGs in these scaling relations will thus require time-dependent, non-equilibrium processes, including bursty star formation histories, stochastic gas accretion, and delayed feedback \citep[e.g.][]{SanchezAlmeida2014_araa,Furlanetto2022,Pallottini2025,Fortune2025}. 
Future work using matched samples across redshift together with detailed comparisons with models and simulations will help determine how burst duty cycles, gas fractions, and selection effects shape the observed evolution of these relations.

\subsection{The low metallicity and high-ionization of EELGs}

The DESI EELG sample allows the identification of physically distinct subsamples associated with extreme nebular conditions. 
These include galaxies with detectable auroral lines such as [O\,{\sc iii}]$\lambda4363$, indicative of low metallicity \citep[e.g.][]{SanchezAlmeida2016,EPM2021}, systems showing high-ionization emission lines such as He\,{\sc ii}$\lambda4686$ or [Ne\,{\sc v}]$\lambda3426$, tracing very hard radiation fields produced by massive stars, shocks, or AGN activity \citep{Kehrig2015,Cleri2023,Mingozzi2025,delValle-Espinosa2025}, and galaxies displaying Mg\,{\sc ii} emission, often associated with enhanced escape of Ly$\alpha$ and Lyman-continuum photons \citep{Henry2018,Chisholm2022,Xu2022b,Leclercq2024}. 
Such properties make EELGs valuable laboratories for investigating the ISM conditions that regulate ionizing-photon escape and thus provide insight into the processes driving cosmic reionization under low-metallicity and high-ionization conditions rarely accessible in normal galaxy samples \citep[e.g.][]{Jaskot2025}.

A particularly informative diagnostic is shown in Fig.~\ref{fig:O32-R23}, where DESI EELGs occupy a well-defined high-ionization sequence in the $O_{32}$-$R_{23}$ plane, clearly offset toward larger values relative to normal local star-forming galaxies. 
Most objects lie at $O_{32}>1$, and a clear trend toward increasing $O_{32}$ with larger nebular equivalent width is observed, consistent with progressively younger and more intense starburst episodes in low-metallicity environments.

The most extreme systems, with $O_{32}>5$ and EW([O\,{\sc iii}])$\gtrsim500$\,\AA, correspond to the regime populated by confirmed low-redshift LyC emitters \citep[e.g.][]{Izotov2018a,Flury2022b}. 
Although only a small fraction of DESI EELGs reach such extreme ionization conditions, their locus partially overlaps with that defined by metal-poor galaxies at $z\sim3$-9. 
This suggests that the physical conditions commonly observed in reionization-epoch galaxies can already be found in rare low-redshift systems.

Interestingly, several DESI EELGs show spectra dominated by high-ionization lines, yielding only lower limits in $O_{32}$ and $R_{23}$. 
These objects challenge standard ionization-bounded photoionization models \citep[e.g.][]{Cullen2025} and may represent promising candidates for density-bounded nebulae with significant LyC leakage. 
Because these galaxies are typically fainter and less massive than previously known LyC emitters \citep{Izotov2018a,Flury2022a}, they provide an opportunity to probe the low-luminosity regime of ionizing-photon escape at low redshift, largely unexplored in SDSS-based samples.

The size of this extreme subsample is expected to increase significantly with future DESI data releases, offering a statistically powerful benchmark for comparison with reionization-epoch galaxies observed with \textit{JWST}.

\section{Conclusions and future extensions}
\label{sec:summary}

In this work we implemented a modified version of the ASK spectral classification method \citep{SA2010}, based on k-means clustering, to efficiently identify extreme emission-line galaxies (EELGs) among minor spectral classes and outliers in the DESI-EDR galaxy sample. 
The ASK approach relies on the global spectral shape defined by selected continuum and emission-line windows and is designed as a fast pre-selection tool to isolate rare objects within massive spectroscopic datasets prior to detailed spectroscopic measurements and physical modelling.

Using this strategy, we constructed and publicly release a large catalog of compact EELGs including emission-line measurements, physical properties from SED fitting, and gas-phase metallicities. 
A first statistical characterization of the sample yields the following main results.

\begin{itemize}
    \item From a parent sample of 820\,240 galaxy spectra in DESI-EDR, we identify 16\,242 EELGs at $z\lesssim0.96$ with extreme nebular equivalent widths (EW$_{5007}\sim100$-2500\,\AA\ and EW$_{\hb}\sim20$-500\,\AA), representing $\sim2\%$ of the parent population. 
    This confirms the rarity of such extreme starbursts in the low-redshift Universe while providing the largest homogeneous spectroscopic sample assembled to date (Sect.~\ref{sec:data}).

    \item Emission-line diagnostic diagrams reveal that $\sim95\%$ of EELGs are compact star-forming systems characterized by high excitation conditions and low metallicities, while $\sim5\%$ show harder spectra and broader emission lines consistent with AGN activity. 
    Composite spectra further indicate that extreme nebular emission often coexists with underlying evolved stellar populations, suggesting recurrent burst episodes rather than purely young systems (Sect.~\ref{sec:diagnostics}).

    \item SED modelling confirms that EELGs typically exhibit bursty star-formation histories, low dust attenuation, and very high specific star-formation rates. 
    In the SFR-$M_{\star}$ plane, they define a tight sequence well above the local star-forming main sequence and broadly consistent with the normalization observed at $z>2$ with \textit{JWST}. The most extreme systems with EW(H$\beta$+[O\,{\sc iii}])$>1000$\,\AA\ lie close to the maximum sSFR allowed by current models at fixed stellar mass ($\sim$100 Gyr$^{-1}$), indicating galaxy-wide young starbursts in small low-mass systems (Sect.~\ref{sec:SFR}).

    \item Direct-method metallicity measurements based on [O\,{\sc iii}]$\lambda4363$ detections reveal a wide range of physical conditions, with electron densities $N_e\sim10$-2000\,cm$^{-3}$ and temperatures up to $\sim3\times10^{4}$\,K. 
    Gas-phase metallicities span $12+\log(\mathrm{O/H})\sim6.9$-8.5, including a non-negligible fraction of extremely metal-poor galaxies ($\lesssim0.1\,Z_{\odot}$) and a few outliers reaching ionization conditions comparable to the most extreme reionization-era galaxies identified with \textit{JWST} (Sect.~\ref{sec:oxygen}; Fig.~\ref{fig:histogram}).

    \item The mass-metallicity relation defined by EELGs is shallow and significantly offset toward lower metallicities relative to normal star-forming galaxies. 
    Its slope and normalization closely match recent direct-method measurements at $z\sim3$-10 from \textit{JWST}. 
    Projecting the sample along the FMR parameter $\mu_{0.65}$ does not reduce the large intrinsic metallicity scatter ($\sim1$ dex), indicating strong departures from equilibrium growth and suggesting that stochastic gas inflow and bursty star formation dominate the baryon cycle in these systems (Sect.~\ref{sec:MZR}; Fig.~\ref{fig:MZR}).

    \item In the $O_{32}$-$R_{23}$ diagram, EELGs populate a high-ionization sequence clearly offset from the bulk of local star-forming galaxies. 
    The most extreme objects reach $O_{32}>5$, overlapping with confirmed low-redshift Lyman-continuum emitters and metal-poor galaxies at $z\sim3$-9. 
    These systems represent promising candidates for density-bounded nebulae and provide valuable low-redshift analogs of galaxies driving cosmic reionization (Sect.~\ref{sec:discussion}; Fig.~\ref{fig:O32-R23}).
 
\end{itemize}

Overall, the DESI EELG sample provides compelling evidence that extreme emission-line selection efficiently isolates galaxies undergoing short-lived, highly non-equilibrium phases of evolution. 
These galaxies appear to be building a significant fraction of their stellar mass in rapid starburst episodes under low-metallicity and high-ionization conditions similar to those observed in the early Universe. 
As such, they constitute key laboratories to study the physical mechanisms regulating star formation, feedback, and ionizing-photon escape across cosmic time.

Finally, this work demonstrates the potential of ASK-based spectral classification as a scalable and efficient strategy to identify rare populations in current and future spectroscopic surveys. 
The extension of this methodology to forthcoming DESI data releases, which will increase sample sizes by orders of magnitude, together with its integration into machine-learning pipelines will enable systematic searches for extreme galaxies and other spectral outliers in the era of massive spectroscopic datasets.

\begin{acknowledgements}
We thank P. Papaderos, S. Oey, and A. Lumbreras for useful discussions. RA and AHC acknowledge support of Grant PID2023-147386NB-I00 funded by MICIU/AEI/10.13039/501100011033 and by ERDF/EU. JAFO and AH acknowledge financial support by the Spanish Ministry of Science and Innovation (MCIN/AEI/10.13039/501100011033), by ``ERDF A way of making Europe'' and by ``European Union NextGenerationEU/PRTR'' through the grants PID2021-124918NB-C44 and CNS2023-145339; MCIN and the European Union -- NextGenerationEU through the Recovery and Resilience Facility project ICTS-MRR-2021-03-CEFCA. 
AGA, JMV and EPM acknowledge support from the Severo Ochoa award to the IAA-CSIC  CEX2021-001131-S, and from the grants PID2022-136598NBC32 “Estallidos8” and PID2019-107408GB-C44 funded by MICIU/AEI/10.
13039/501100011033. JSA acknowledges financial support from the Spanish Ministry of Science and Innovation, project PID2022-136598NB-C31 (ESTALLIDOS8) and from the EU UNDARK project (project number 101159929). LB, SS, and RA acknowledge support from ANID Fondecyt 1202007. 

\end{acknowledgements}

%
   \bibliographystyle{aa} 
   \bibliography{biblio} 
%



\begin{appendix} 

\section{Additional material on ASK classes}
\label{app:ask}

\subsection{ASK workflow}

\begin{figure}[!t]
    \centering
\includegraphics[width=0.50\textwidth]{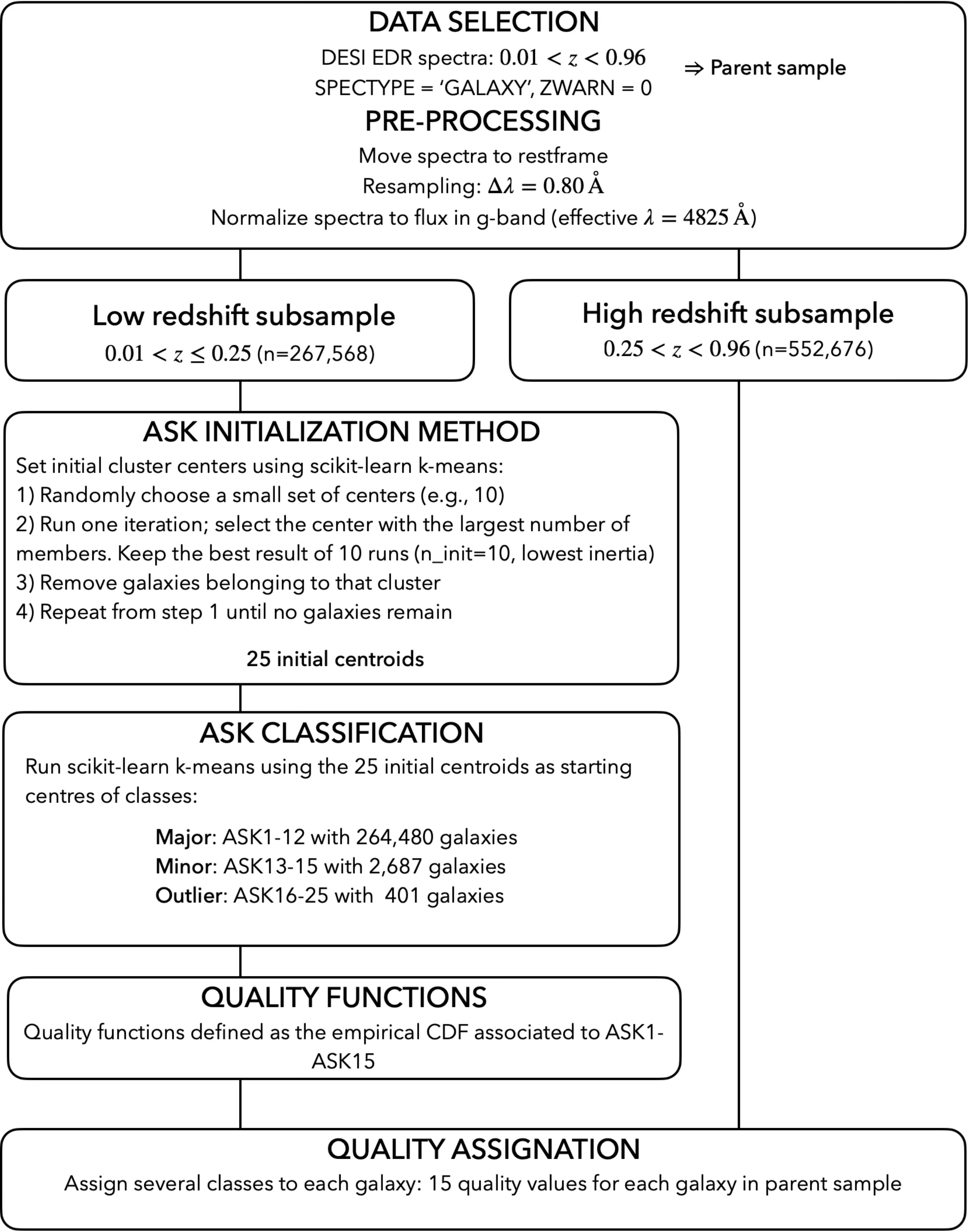}
    \caption{Diagram showing the steps of the ASK method. See section 3 for further details.}
    \label{fig:proc-steps}
\end{figure}

The workflow of the ASK classification applied to DESI EDR spectra is illustrated in Fig.~\ref{fig:proc-steps}. 
Data selection is restricted to galaxies (\texttt{SPECTYPE = GALAXY}, \texttt{ZWARN = 0}) in the redshift range $0.01 < z < 0.96$. 
Pre-processing includes shifting spectra to the rest frame, resampling to $\Delta\lambda = 0.80$\,\AA, and normalizing to the flux in the $g$-band ($\lambda_{\mathrm{eff}} = 4825$\,\AA). 
The parent sample is then divided into two subsets: a low-redshift subsample ($0.01 < z \leq 0.25$, 267,568 spectra) and a high-redshift subsample ($0.25 < z < 0.96$, 552,676 spectra). 

For the low-redshift subset, initial cluster centers are defined using a modified \texttt{scikit-learn} k-means procedure:  
(1) randomly select a small set of centers (e.g., 10);  
(2) run one iteration and select the center with the largest number of members, keeping the best of 10 runs ($n_{\mathrm{init}}=10$, lowest inertia);  
(3) remove galaxies belonging to that cluster;  
(4) repeat until no galaxies remain.  
This procedure yields 25 initial centroids. 

The ASK classification is then performed with these centroids as starting centers, producing three groups of classes: Major (ASK1--12; 264,480 galaxies), Minor (ASK13--15; 2,687 galaxies), and Outliers (ASK16--25; 401 galaxies). 
Quality functions are defined as empirical cumulative distribution functions (CDFs) for ASK1--ASK15, while classes ASK16--ASK25 lack quality functions due to insufficient populations. 
Finally, quality assignation provides each galaxy in the parent sample with 15 quality values.

Fig. \ref{fig:q_ratio} shows histograms of the ratio between the best and second-best quality values for the Main and Minor ASK classes (ASK0–ASK14). 
This ratio provides an empirical measure of how well-separated each class is in the high-dimensional classification space \cite[fig. 8]{SA2010}. In our case, four classes (e.g., ASK10, ASK12) display distributions that peak at ratios significantly smaller than one, indicating the presence of well-defined clusters. Conversely, most other classes, such as ASK0 and ASK2, show distributions that peak near one, reflecting similar best and second-best quality values and suggesting a high fraction of borderline galaxies.

\begin{figure*}
    \centering
\includegraphics[width=0.99\textwidth]{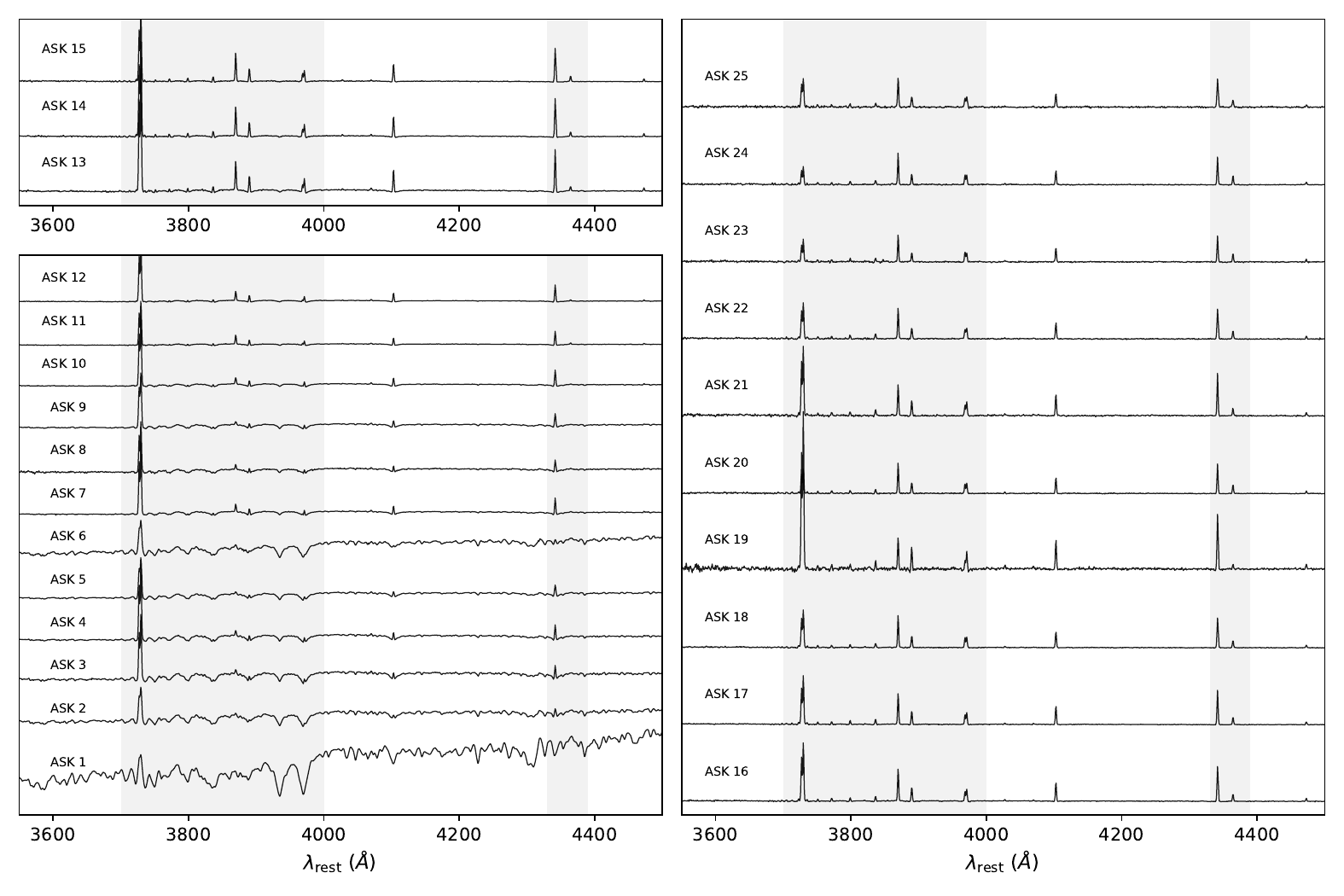}
    \caption{Zoom-in version of  Fig.~\ref{fig:clusters} to show spectral feature in the bluest bands. We note the emergence of stellar absortion features in the major ASK clases (bottom-left panel) and the rise of faint nebular emmision lines in the minor ASK clases (top-right panel) and outliers (left panel).}
    \label{fig:outliers}
\end{figure*}

\begin{table}
\caption{\label{tab:windows} Spectral windows selected for ASK classification in DESI.}
\centering
\begin{tabular}{cc}
\hline\hline
$\lambda$ Bands &  Comments \\
\hline 
(3700.0, 4000.0) & Balmer break, [\oii], [\neiii]  \\
(4330, 4390) & \hg, [\oiii]  \\
(4620, 4780) & WR bump, \heii  \\
(4830, 5040) & \hb, [\oiii] doublet  \\
(5880, 5910) & NaD, \hei   \\
(6290, 6380) & TiO$_2$, [\oi], [\siii]  \\
(6530, 6750) & \ha, [\nii], [\sii]  \\
(7050, 7155) & TiO band  \\
(7270, 7360) & [\oii]  \\
\hline
\hline
\end{tabular}
\tablefoot{Wavelength are in \AA. The comments highlight the main spectral features of interest in the corresponding band.}
\end{table}

\begin{table}
\caption{\label{tab:ASK-EW} Emission-line measurements for the ASK centroid spectra.}
\centering
\begin{tabular}{cccc}
\hline\hline
ASK centroid& $EW_{\hb}$& $EW_{5007}$& $EW_{\ha}$\\
\hline 
 ASK\,10  &14.5$\pm$0.9&30.1$\pm$0.7&64.0$\pm$1.8\\
 ASK\,11  &18.9$\pm$0.6&67.1$\pm$2.2&78.3$\pm$2.7 \\
 ASK\,12  &23.4$\pm$1.3&67.6$\pm$1.8&102.5$\pm$3.3 \\
\hline
 ASK\,13  &33.2$\pm$1.8&121.0$\pm$4.0&146.7$\pm$5.0\\
 ASK\,14  &45.7$\pm$2.6&183.7$\pm$10.7&201.8$\pm$7.5 \\
 ASK\,15  &61.8$\pm$4.3&283.7$\pm$20.6&272.3$\pm$13.1\\
\hline
 ASK\,16  &82.0$\pm$5.9& 428$\pm$42 & 382$\pm$25 \\
 ASK\,17  &113.4$\pm$10.1& 598$\pm$71 & 534$\pm$46 \\
 ASK\,18  &135.6$\pm$16.7& 833$\pm$140 & 677$\pm$98 \\
 ASK\,19  &114.6$\pm$16.8& 352$\pm$83 & 306$\pm$85 \\
 ASK\,20  &161.5$\pm$24.6& 1118$\pm$276 & 775$\pm$182  \\
 ASK\,21  &173.5$\pm$41.1& 715$\pm$126 & 758$\pm$169 \\
 ASK\,22  &199.4$\pm$54.6& 1317$\pm$352 & 982$\pm$306 \\
 ASK\,23  &171.6$\pm$57.7& 1068$\pm$297 & 888$\pm$369 \\
 ASK\,24  &208.9$\pm$60.2& 1401$\pm$339 & 1048$\pm$424 \\
 ASK\,25  &208.5$\pm$83.8& 1240$\pm$359 & 1079$\pm$530  \\
\hline
\end{tabular}
\tablefoot{Equivalent widths are indicated in \AA. Major ASK classes between ASK1 and ASK9 have in all cases lower EW values compared to ASK10-12 (see also Fig.~\ref{fig:hist_ask}). They are not displayed.}
\end{table}

\begin{figure*}[!t]
    \centering
\includegraphics[width=0.80\textwidth]{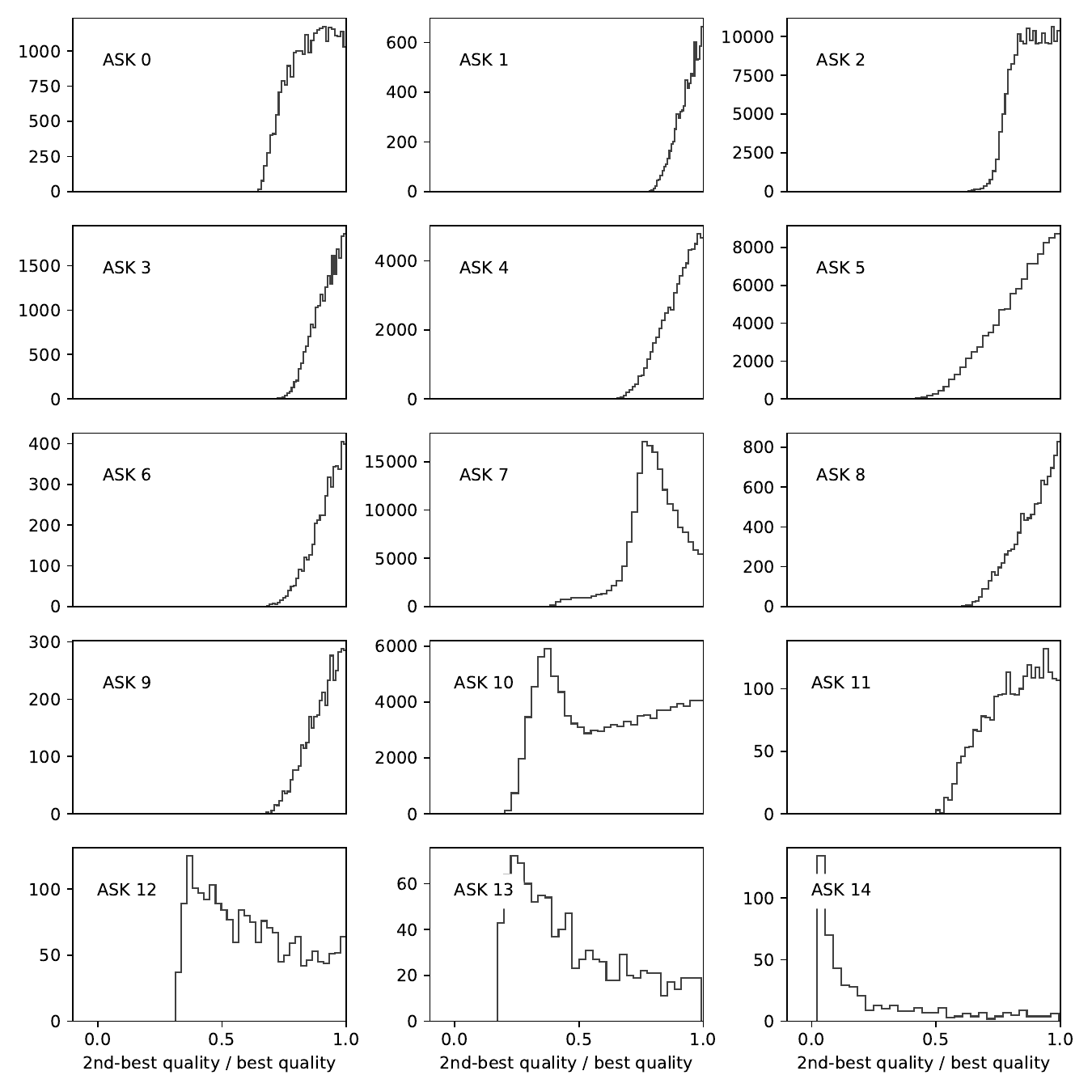}
    \caption{Distribution of the ratio between the best and second-best quality values for galaxies in the MAJOR and MINOR ASK classes. Classes that represent well-defined clusters in the 3104-dimensional classification space are expected to show ratio distributions with peaks significantly smaller than one (e.g., ASK12). Only galaxies with a best quality value exceeding 0.2 are included. }
    \label{fig:q_ratio}
\end{figure*}

\subsection{Composite spectra of DESI EELGs}
\label{app:composite}

\begin{figure*}[!t]
    \centering
\includegraphics[width=0.99\textwidth]{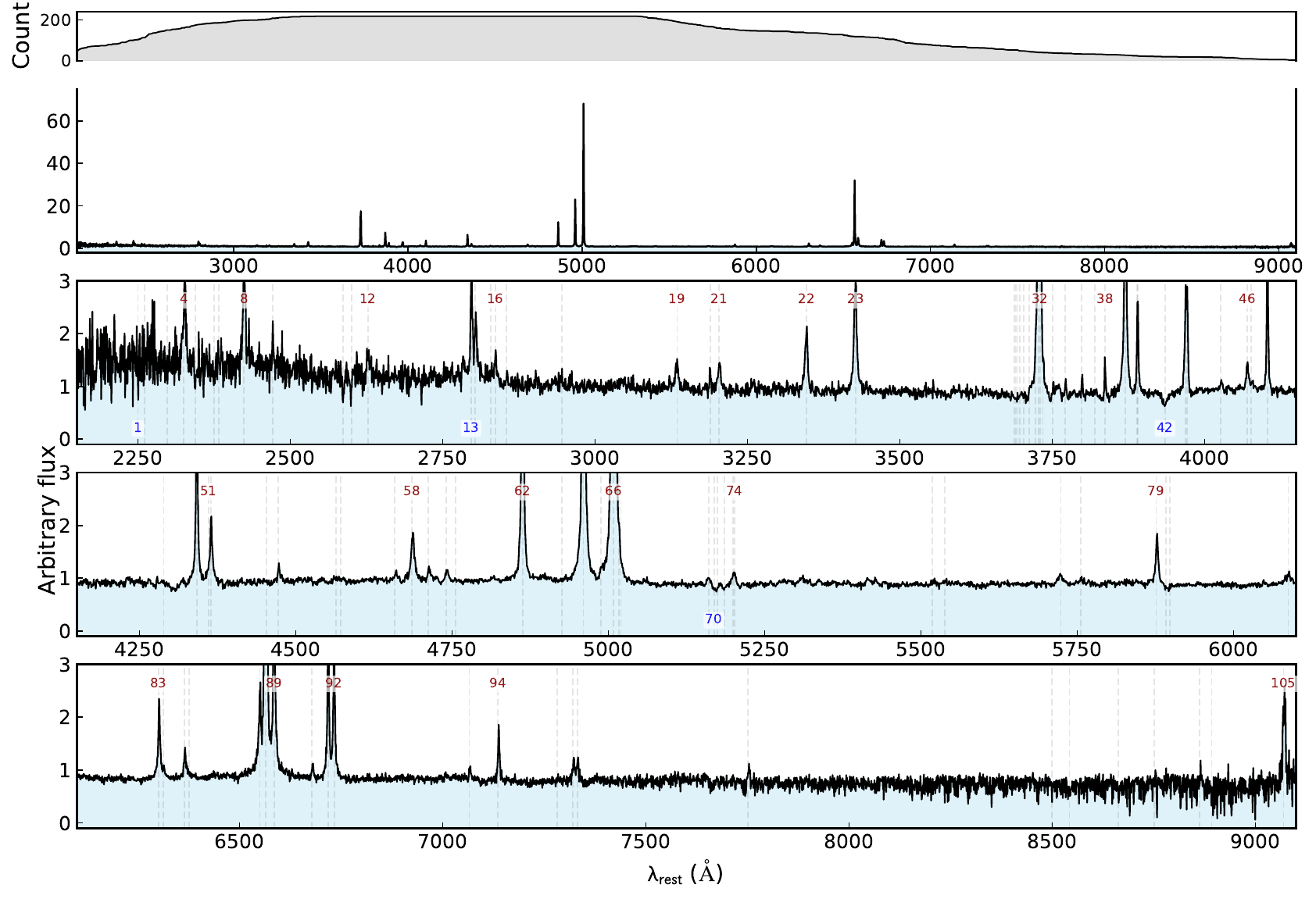}
    \caption{Median stack of our AGN candidates in the k-MENDEL EELG sample, according to emission-line diagnostics (see Section~\ref{sec:diagnostics}). Both emission and absorption line features are indicated by grey dotted lines, and correspond to ions and transitions described in Table~\ref{tab:composite}. }
    \label{fig:agn-stack}
\end{figure*}

We provide composite spectra for star-forming and AGN galaxies obtained from our EELG ASK classification as online machine-readable tables. We also make publicly available a short Python script to read the spectra and display them interactively. While the star-forming EELGs are shown in Figure~\ref{fig:median-stack}, we display in Figure~\ref{fig:agn-stack} the median stack of our AGN candidates in the k-MENDEL EELG sample.  

The composite spectra was obtained using the same procedure as in ASK cluster templates in Fig.~\ref{fig:clusters}, i.e. a median-weighted stacking scheme after resampling the deredshifted spectra to a common $\Delta\lambda$\,0.8\AA\ wavelength step and normalizing the spectra using a spectral continuum window of 20\AA\ centered at 4800\AA. The redshift range of our sample allows us to obtain a full stack with wavelength limits between $\sim$\,2200\AA\ and $\sim$\,9100\AA, thus allowing us to probe numerous emission and absorption features from [\feii]\,2250\AA\ to [\siii]9062\AA, as identified in Table~\ref{tab:composite}. The main purpose of this exercise is to obtain a representative spectrum with enough S/N to identify the faintest features that cannot be detected in single spectra, serving as an atlas of typical EELG that can be used as templates for several case studies. 

\begin{table}
\caption{\label{tab:composite} Emission and absorption lines identified in median stacks of the entire k-MENDEL EELG sample.  The upper panel shows the number of galaxies contributing to each wavelength.}
\centering
\begin{tabular}{lccc}
\hline\hline
Number & Line \texttt{id} & $\lambda_0$ (vacuum) &  Comments \\
\hline 
1 & Fe{\sc ii}	& 2249.9 & absorption, ISM/CGM \\
2 & Fe{\sc ii}  & 2260.8 & absorption, ISM/CGM \\
3 & C{\sc iii}  & 2297.6 & absorption, ISM/CGM \\
4 & C{\sc ii}   & 2326.0 & emission, ISM \\
...  & ... & ...& ... \\
...  & ... & ...& ... \\
100 & Pa11 & 8862.8 & emission H{\sc ii} \\
101 & Fe{\sc ii}]   & 8891.9  & emission ISM \\
102 & [S{\sc iii}]  & 9069.0 & emission, nebular \\ 	
\hline
\hline
\end{tabular}
\tablefoot{Wavelength are in \AA. The comments highlight the main spectral features of interest in the lines and their likely origin. An entire electronic version of this table is available online. }
\end{table}

\section{Catalog description}
\label{app:cat}

In this section, we briefly describe the catalog of EELGs from the k-MENDEL sample. 
The structure of the emission-line catalog is presented in Table~\ref{tab:eelg_catalog_preview}. 
The full tables are available online through CDS and on a dedicated webpage\footnote{\url{www.zenodo...}}, where we will provide updated versions for future DESI public data releases.

Emission-line measurements were obtained with our tool \textsc{LiMe} \citep{LiMe2024}. Gaussian fits, as described in Section~\ref{sec:methods}, provide the line center (in \AA), width (in \AA), and flux (in erg s$^{-1}$ cm$^{-2}$). 
We also report equivalent widths (EWs), computed from the ratio between line flux and local continuum flux. Uncertainties are provided for all measured quantities. 
Each galaxy is identified by its DESI \texttt{TARGETID} and spectroscopic redshift (\texttt{Z}). The catalog also includes the extinction coefficient $c(\hb)$ derived as described in Section~\ref{sec:methods}.

In addition, a second table provides key emission-line ratios and derived physical properties for the k-MENDEL sample. A preview of these quantities is shown in Table~\ref{tab:eelg_catalog_physprop_preview}. 
The catalog is designed to facilitate reproducibility and enable further analysis of the results presented in this work.

\begin{table*}
\caption{Preview of the emission-line catalog for the DESI EELG sample. 
The full catalog is available in electronic form at the CDS.}
\label{tab:eelg_catalog_preview}
\centering
\begin{tabular}{cccccccc}
\hline\hline
TARGETID & RA & DEC & $z$ & $F_{\rm [OII]3726}$ & $eF_{\rm [OII]3726}$ & $F_{\rm [OII]3729}$ & $eF_{\rm [OII]3729}$ \\
 & (deg) & (deg) & \multicolumn{4}{c}{$(10^{-17}$ erg s$^{-1}$ cm$^{-2})$} \\
\hline
39627087803327287 & 60.902848 & -30.208123 & 0.1174 & 140.09 & 10.08 & 188.00 & 14.59 \\
39627087815906772 & 61.706061 & -30.227423 & 0.1589 & 74.34  & 5.15  & 84.67  & 5.32 \\
39627093025235111 & 60.387177 & -30.031876 & 0.0640 & 33.13  & 6.64  & 34.59  & 6.39 \\
39627093037817386 & 61.276922 & -30.117559 & 0.1086 & 90.71  & 6.03  & 116.13 & 7.59 \\
\hline
\end{tabular}
\tablefoot{Only the first rows and columns are shown here for guidance regarding the form and content of the catalog. 
The full table, including measurements for all emission lines and derived quantities, is available in electronic form.}
\end{table*}

\begin{table*}
\caption{Preview of the catalog of phsyical properties for the DESI EELG sample. 
The full catalog is available in electronic form at the CDS.}
\label{tab:eelg_catalog_physprop_preview}
\centering
\begin{tabular}{lcccccccc}
\hline\hline
TARGETID & z & $R_{e}$ & $\log(M_{\star})$ & $\log(SFR_{\rm 10,Balmer})$ & $n_{e}$ & $t_{e}$ & 12$+\log$(O/H) \\
    & & (kpc)  &  ($M_{\odot}$) & ($M_{\odot}$yr$^{-1}$) & (cm$^{-3}$) & (10$^{4}$K) & dex   \\
\hline
49749096923136  & 0.0192 & ...  & 6.47 & -1.70 & 26 & 1.87 & 7.55 \\
103614278270984 & 0.4397 & ...  & 7.73 &  0.06 & 99 & 1.56 & 7.84 \\
 ...            & ...    & ...  & ... & ... & ... & ... & ... \\
103621320507411 & 0.1962 & 1.40 & 8.10 & -0.23 & 34 & 1.62 & 7.83 \\
103625112158254 & 0.3855 & ...  & 7.30 & -0.60 & 125 & 1.74 & 7.75 \\
 ...            & ...    & ...  & ... & ... & ... & ... & ... \\
\hline
\end{tabular}
\tablefoot{Only the first rows and columns are shown here for guidance regarding the form and content of the catalog. 
The full table, including derived physical properties, is available in electronic form.}
\end{table*}

\begin{table}[h]
		
		\tiny
		\centering
		\caption{\texttt{CIGALE} Parameters}
		\label{cigale_tab}
		\begin{tabular}{cc}
			\toprule
			\toprule
			\textbf{Stellar Parameters} &  \\ \midrule
			$\tau_{\mathrm{main}}$ {[}Myr{]} & 5, 10, 20, 50 \\ \rule{0pt}{2.3ex}
			$\tau_{\mathrm{burst}}$ {[}Myr{]}      &  0.5, 0.8, 1, 3 \\ \rule{0pt}{2.3ex}
			Age main {[}Myr{]} & 30, 50, 100, 200, 500        \\ \rule{0pt}{2.3ex}
			burst\_age {[}Myr{]} & 1, 3, 6, 10, 12 \\ \rule{0pt}{2.3ex}
			f\_burst& 0.1, 0.15, 0.25, 0.3, 0.5 \\
			\midrule
			\textbf{Charlot \& Bruzual (2019)} &  \\ \midrule 
			IMF & Chabrier \\ \rule{0pt}{2.3ex}
			metallicity & 0.0001, 0.001, 0.004, 0.008\\ \rule{0pt}{2.3ex}
			Upper IMF limit [M$_\odot$] & 100\\

			\midrule 
			\textbf{Nebular parameters} &  \\ \midrule 
			z\_gas & 0.0001, 0.001, 0.004, 0.008   \\ \rule{0pt}{2.3ex}
			$\log{U}$  & -3.5, -3.0, -2.5, -2.0, -1.5   \\ \rule{0pt}{2.3ex}
			f$_{esc}$, f$_{dust}$ & 0\\
			
			\midrule
			\textbf{Extinction parameters} &  \\ \midrule
			$E(B-V)_\mathrm{young}$  & 0.1, 0.2, 0.3, 0.4  \\ \rule{0pt}{2.3ex}
			$E(B-V)_\mathrm{old\_factor}$    & 0.44, 1     \\ 
			\noalign{\smallskip}
			
			\bottomrule
		\end{tabular}
	\end{table}

\subsection{Possible caveats}

We identify possible caveats in the dataset that are worth mentioning as they might affect spectrophotometry with DESI EDR and, in particular, our sample of EELGs. The first of such caveats is related to the global spectrophotometry of DESI EDR spectra. We identified a small fraction of galaxies for which the relative flux calibration of some of the blue, green and red arms, is inconsistent with the other two, resulting in flux "jumps" in the spectra. This generally affects the red arm, resulting in artificial red $g-i$ (or $r-z$) colours, reddens the visible SED slope, and therefore affects stellar masses and other SED parameters derived by CIGALE (such as stellar ages and dust extinction). This caveat was partly addressed by applying a cross-match with \citet{Lan2023}, who visually identified a large number of such objects, but not all of them. In this work, we decided to keep these objects in our k-MENDEL EELG sample but we caution the readers to check carefully rest-frame colors if spectrophotometric measurements are needed from our catalog. It is also worth mentioning that our LiMe measurements for emission lines are not affected by such systematics because LiMe fits and subtracts the \textit{local } continuum instead of a global one, as in the available VACs \citep[e.g.][]{Zou2024}. 

Another identifiable caveat in our analysis concerns the DESI EDR flux calibration in the blue arm, which, in some galaxies, shows a strong artificial decrease in continuum flux. This is due to imperfect flux calibration and is expected to be improved with future data releases. For this reason, the analysis and discussions presented in this paper are restricted to curated subsamples of EELGs at $z>0.01$, for which we can ensure reliable key emission line ratios.  
Moreover, the presence of double-peaked emission-line spectra, which can be real (e.g., \texttt{TARGETID\,39632930582039713}), due to residuals from an imperfect subtraction of strong telluric features (e.g., \texttt{TARGETID\,39627136830541283}), or both (e.g., \texttt{TARGETID\,39633315484926470}), might affect emission-line measurements in our catalog, especially at wavelengths $\lambda \gtrsim$\,9300\AA. 
This is especially true for [\oiii]4959,5007 for galaxies at $z\gtrsim$\,0.85 and \ha\ lines for galaxies at $z\gtrsim$\,0.40. 

While we keep all the originally selected k-MENDEL EELG in the master catalog, we decided to remove those subsets that are potentially affected by any of the above systematics in the analysis presented in this paper (see the text for details). Those galaxies will be considered for future works using more advanced data releases.  We stress that removing such subsets does not affect any of our conclusions.

\end{appendix}
\end{document}